\documentclass[article,nojss,shortnames]{jss}

\usepackage[]{graphicx}
\usepackage[]{xcolor}
\makeatletter
\def\maxwidth{ %
  \ifdim\Gin@nat@width>\linewidth
    \linewidth
  \else
    \Gin@nat@width
  \fi
}
\makeatother

\usepackage[utf8]{inputenc}
\usepackage{thumbpdf,lmodern}
\usepackage{xspace}

\usepackage{amsfonts,amstext,amsmath,amssymb}
\usepackage{nicefrac}
\usepackage{accents} 

\usepackage[sectionbib]{bibunits}
\defaultbibliographystyle{jss} 
\defaultbibliography{mlt,surv,packages}

\usepackage[title]{appendix}

\usepackage{lineno} 

\usepackage{booktabs}
\makeatletter
\g@addto@macro{\table}{\centering}
\makeatother
\usepackage{floatrow} 
\usepackage[final]{changes}

\usepackage{verbatim}
\usepackage{alltt}
\usepackage{paralist}

\renewcommand{\code}[1]{\texttt{#1}}
\newcommand{\var}[1]{\code{#1}}
\newcommand{\cmd}[1]{\code{#1()}}

\newcommand{\sMEV}[1]{\exp\left\{-\exp\left[#1\right]\right\}}
\newcommand{\rS}{S}
\newcommand{\rs}{s}

\newcommand{\escaleparm}{\gamma}
\newcommand{\rF}{A}
\newcommand{\rf}{a}

\newcommand{\rC}{C}
\newcommand{\rc}{c}
\newcommand{\rW}{W}
\newcommand{\rw}{w}
\newcommand{\sw}{S_\rw}
\newcommand{\sYw}{S_\rw}
\newcommand{\sCw}{G_\rw}
\newcommand{\indep}{\perp \!\!\! \perp}
\newcommand{\rA}{\text{Age}}
\newcommand{\ra}{\text{age}}

\newcommand{\lparm}{\xi}
\newcommand{\mparm}{\lparm}

\newcommand{\haz}{\lambda}
\newcommand{\Haz}{\Lambda}
\newcommand{\haznul}{\haz_0}
\newcommand{\hazone}{\haz_1}
\newcommand{\Haznul}{\Haz_0}
\newcommand{\hatHaznul}{\hat{\Haz}_0}
\newcommand{\Hazone}{\Haz_1}

\newcommand{\snul}{S_{0}}
\newcommand{\sone}{S_{1}}
\newcommand{\dnul}{f_{0}}

\newcommand{\pnul}{F_{0}}

\newcommand{\rZ}{Z}
\newcommand{\rY}{T}

\newcommand{\rz}{z}
\newcommand{\ry}{t}
\newcommand{\rx}{\xvec}

\newcommand{\dZ}{f}

\newcommand{\h}{h}

\newcommand{\basisy}{\avec}
\newcommand{\bern}[1]{\avec_{\text{Bs},#1}}

\newcommand{\parm}{\varthetavec}
\newcommand{\eparm}{\vartheta}
\newcommand{\dimparm}{P}

\newcommand{\eshiftparm}{\beta}

\newcommand{\ie}{\textit{i.e.,}}

\renewcommand{\Prob}{\mathbb{P}}
\newcommand{\Ex}{\mathbb{E}}
\newcommand{\RR}{\mathbb{R}}

\usepackage{dsfont}

 \DeclareMathOperator{\logit}{logit}

 \DeclareMathOperator{\MEV}{MEV}

\def \avec {\text{\boldmath$a$}}

    \def \mR {\text{\boldmath$R$}}

\def \xvec {\text{\boldmath$x$}}

\def \gammavec        {\text{\boldmath$\gamma$}}

\def \varthetavec     {\text{\boldmath$\vartheta$}}

\newcommand{\ubar}[1]{\underaccent{\bar}{#1}}

\definecolor{Red}{rgb}{0.5,0,0}
\definecolor{Blue}{rgb}{0,0,0.5}

\title{Smooth Transformation Models for Survival Analysis: A Tutorial Using \proglang{R}}
\Shorttitle{Smooth Transformation Models for Survival Analysis: A Tutorial Using \proglang{R}}
\Plaintitle{Smooth Transformation Models for Survival Analysis: A Tutorial Using R}
 
\author{Sandra Siegfried \\ Universit\"at Z\"urich \And B\'alint Tam\'asi \\ Universit\"at Z\"urich \And Torsten Hothorn \\ Universit\"at Z\"urich}
\Plainauthor{Siegfried, Tam\'asi and Hothorn}
 

\Plainkeywords{non-proportional hazards, dependent censoring, clustered observations, personalised medicine, survival trees, R}
 
\Address{
Sandra Siegfried, B\'alint Tam\'asi, and Torsten Hothorn\\
Institut f\"ur Epidemiologie, Biostatistik und Pr\"avention \\
Universit\"at Z\"urich \\
Hirschengraben 84, CH-8001 Z\"urich, Switzerland \\
\texttt{Siegfried.Sandra@protonmail.ch}, \texttt{Torsten.Hothorn@uzh.ch} \\
}
 
\Abstract{
Over the last five
decades, we have seen strong methodological advances in survival analysis, using parametric methods and, more prominently, methods based on non-/semi-parametric estimation. As the methodological landscape continues
to evolve, the task of navigating through the multitude of methods and
identifying available software resources is becoming increasingly challenging -- especially in more complex scenarios, 
such as when dealing with interval-censored or clustered survival data,
non-proportional hazards, or dependent censoring.

This tutorial explores the potential of using the framework of smooth transformation
models for survival analysis in the \proglang{R} system for statistical computing.
This framework provides a unified maximum-likelihood approach that covers a wide
range of survival models, including well-established ones such as
the Weibull model and a fully parametric version of the famous Cox
proportional hazards model, and various extensions for more complex scenarios. 
We explore models for non-proportional/crossing hazards, dependent censoring, clustered observations
and extensions towards personalised medicine within this framework.

Using survival data from a two-arm randomised controlled
trial on rectal cancer therapy, we demonstrate how survival analysis tasks can be seamlessly
navigated in \proglang{R} within this framework using
the implementation provided by the \pkg{tram} package, and
few related packages.
 }
\IfFileExists{upquote.sty}{\usepackage{upquote}}{}
\begin{document}

\begin{bibunit}

\vspace{-.2cm}

\section{Introduction}

In ``parametric'' survival analysis, researchers commonly rely on the Weibull
model or alternative accelerated failure time models.  To achieve
greater flexibility and overcome the strict distributional assumptions
underlying these models, researchers often need to turn to
non-/semi-parametric methods to analyse their survival data.  When it comes
to non-/semi-parametric approaches, however, overcoming issues tied to
interval-censored or truncated observations can prove challenging due
to their limited availability in standard software implementations.

Furthermore, when aiming to fit models that handle crossing or non-proportional
hazards, clustered observations, or dependent censoring, researchers often
find themselves navigating a complex landscape of diverse software
implementations.  Even the same models \added{can be difficult to compare}\deleted{are sometimes hard to compare} across
different implementations, because different parametrisations,
estimation strategies or optimisation procedures are used. 
This becomes even more challenging when \added{comparing} different
models computed from different packages \deleted{shall be compared} -- emphasising
the benefit of a unified framework that facilitates seamless transitions
between different models and is based on a common core infrastructure 
for model parametrisation, estimation strategy, and optimisation procedure.

To tackle these challenges, researchers may explore the
potential of the \pkg{tram} add-on package \citep{pkg:tram} in \proglang{R}
\citep{R}, which offers a flexible framework for survival analysis.  The
\pkg{tram} package implements a user-friendly interface to smooth
transformation models \citep{Hothorn_Moest_Buehlmann_2017}, which encompass
a range of classical survival models including accelerated failure models
and the Cox proportional hazards model, as well as many useful extensions
and novel model formulations.  The package can be easily installed from the
Comprehensive \proglang{R} Archive Network (CRAN):
\begin{Schunk}
\begin{Sinput}
R> install.packages("tram")
R> library("tram")
\end{Sinput}
\end{Schunk}

All models are implemented in a fully parametric fashion, allowing for
straightforward maximum likelihood inference.  The estimation of these
models is facilitated by the core infrastructure package \pkg{mlt}
\citep{Hothorn_2018,pkg:mlt} which provides a unified maximum likelihood
framework for general transformation models
\citep{Hothorn_Moest_Buehlmann_2017}.  We further leverage two add-on
packages from this family of packages for transformation modelling: The
\pkg{tramME} package \citep{Tamasi_2025,pkg:tramME} implementing mixed-effects and
non-linear additive effects for smooth transformation models, and the
\pkg{trtf} package \citep{pkg:trtf} for estimating tree-based
survival models.

In this tutorial, we will explore a variety of models commonly utilised in
survival analysis.  The focus of this tutorial lies on the practical
implementation and interpretation of these models within the framework of
smooth transformation models, rather than on theoretical aspects.  Our
objective is to provide users with a practical understanding of how to apply
these models using \proglang{R}.  Through an application to data from a
randomised trial on rectal cancer therapy, we showcase how users can
seamlessly conduct their survival analysis tasks without the need to
navigate through a multitude of packages in \proglang{R}.

In Section~\ref{sec:iObs}, we discuss models for independent observations. 
We start with the well-known Weibull model, and then, to introduce more
flexibility and overcome the log-linear log-cumulative hazard assumption
inherent to the Weibull model, we explore a fully parametric version of the
Cox model.  We further discuss the estimation of stratified log-cumulative
hazard functions to account for baseline risk variations across patient
strata.  Moving beyond the assumption of proportional hazards, we showcase
models that challenge this assumption.  We discuss a location-scale version
of the Cox model, accommodating scenarios with non-proportional/crossing
hazards, and models estimating time-varying treatment effects.

Addressing scenarios where the assumption of independent censoring might not
be justified, we discuss a copula proportional hazards model, that
accommodates dependent censoring (Section~\ref{sec:dCens}).  For clustered
observations we employ mixed-effects Cox models and alternative models
featuring marginally interpretable hazard ratios in Section~\ref{sec:dObs}. 
Our tutorial also explores the domain of personalised medicine, presenting
models that incorporate covariate-dependent treatment effects and survival
trees (Section~\ref{sec:PM}).  In Section~\ref{sec:ext}, we explore further
extensions, including topics like frailty models, model estimation using the
non-parametric likelihood, covariate adjustment and the potential of using these models for
sample size estimation of new trials.

This tutorial is composed of the main text, which introduces the models
and very briefly shows how to estimate them using smooth transformation models
in \proglang{R}. In addition,  we present head-to-head
comparisons of user-interfaces and numerical results obtained from alternative
packages available in the \proglang{R} universe in Supplementary Material~\ref{sec:supp}.
Both parts come with
much more detailed \proglang{R} code for exploring fitted models
(for example, plotting model terms, computing confidence intervals, or performing tests),
which can be explored in the corresponding demo:
\begin{Schunk}
\begin{Sinput}
R> demo("survtram", package = "tram")
\end{Sinput}
\end{Schunk}
In our Supplementary Material~\ref{sec:supp}, we conduct a thorough comparison of a
subset of the models discussed here which can be estimated using alternative
implementations (in total 13~established CRAN packages) and corresponding results obtained with smooth transformation models from
\pkg{tram} and \pkg{tramME}.  This quality assurance task not only helped to validate the
implementation in \pkg{tram} and \pkg{tramME} but also led to the identification of
problematic special cases and, in some instances, practically relevant
discrepancies between different package implementations of the very same model. 
Moreover, Supplementary Material~\ref{sec:supp} presents the different user-interfaces
of the different packages side-by-side, such that it becomes simpler to
estimate and compare relatively complex models across independent
implementations.  For the analysis of future survival trials, an assessment
of the agreement of such estimates, standard errors, and possibly other
model aspects can help to increase trust in reported numerical results and
conclusions derived therefrom.

\section{Application} \label{sec:app}

In our tutorial, we will work with data from the CAO/ARO/AIO-04 two-arm
randomised controlled trial
\citep{Roedel_Graeven_Fietkau_2015}, a phase~3
study that aimed to compare an experimental regimen with the previously
established treatment regimen (control) for locally advanced rectal cancer.  In this
experimental regimen, Oxaliplatin was added to the control treatment of preoperative
Fluorouracil-based chemoradiotherapy and postoperative chemotherapy of
locally advanced rectal cancer.

The trial was conducted across
88~centers and included a cohort of
1'236 patients.  The patients were randomly allocated
to the two treatment arms $\rW \in \{0, 1\}$, receiving the experimental
treatment of Oxaliplatin added to Fluorouracil-based preoperative
chemoradiotherapy and postoperative chemotherapy (5-FU + Oxaliplatin, $\rW =
1$) or the control treatment using Fluorouracil only (5-FU, $\rW
= 0$).  Treatment allocation was performed using block-randomisation
stratified by study center $j = 1, \dots,
88$ and the stratum~$\rs$, which is
defined by four categories consisting of a combination of clinical
N~category, \ie~lymph node involvement
(cN0 vs cN+), and clinical
T~category \ie~tumor grading (cT1-3 vs cT4).  The distribution of patients in the two treatment arms across
strata is shown in Table~\ref{tab:strat}.

\begin{table}
\caption{Number of patients in each treatment arm stratified by the combination of
clinical N~and T~category.} \label{tab:strat}
\centering
\small
%

\begin{tabular}{lrr}
  \toprule
 & 5-FU & 5-FU + Oxaliplatin \\ 
  \midrule
cT1-3\phantom{4} : \phantom{-}cN0 & 163 & 156 \\ 
  cT4\phantom{1-3} : \phantom{-}cN0 & 8 & 7 \\ 
  cT1-3\phantom{4} : \phantom{-}cN+ & 411 & 417 \\ 
  cT4\phantom{1-3} : \phantom{-}cN+ & 41 & 33 \\ 
   \midrule
Total & 623 & 613 \\ 
   \bottomrule
\end{tabular}

\end{table}

\begin{figure}[t!]
\begin{Schunk}

{\centering \includegraphics{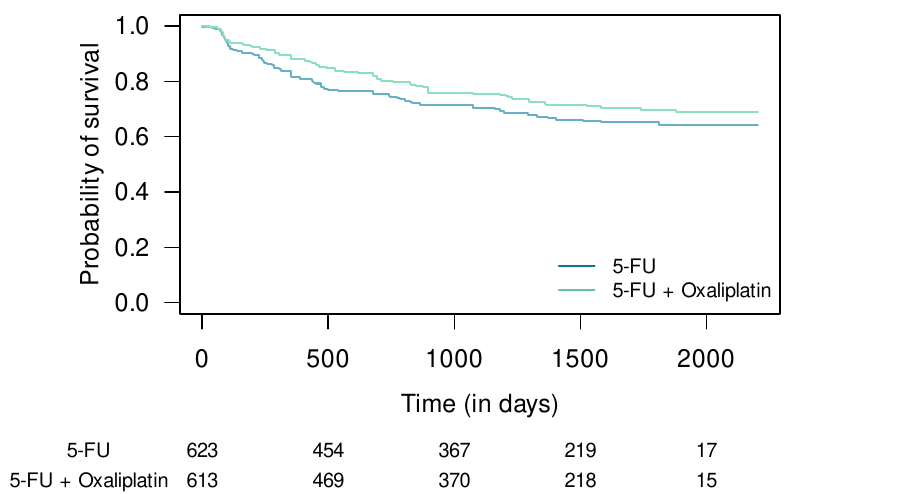} 

}

\end{Schunk}
\caption{Disease-free survival. The survivor functions of the two treatment arms estimated
by the non-parametric Turnbull method are shown together with the number at risk table.}
\label{fig:DFS}
\end{figure}

The primary endpoint is disease-free survival, defined
as the time $\rY \in \RR^{+}$ between randomisation and non-radical surgery
of the primary tumor (R2 resection), loco-regional recurrence after R0/1
resection, metastatic disease or progression, or death from any cause --
whichever occurred first.  The observed times encompass a mix of exact
observations $\ry$ for time to death or incomplete removal of the primary
tumor, interval-censored observations $\ry \in (\ubar{\ry}, \bar{\ry}]$ for the 
time span from the previous follow-up $\ubar{\ry}$ to the time-point of detecting local or distant
metastases $\bar{\ry}$, and right-censored observations $\ry \in (\ry,
\infty)$ corresponding to the end of the follow-up period or instances where
patients were lost to follow-up. The survivor curves of the primary
endpoint (disease-free survival) estimated by the non-parametric Turnbull
estimator \citep{Turnbull_1976} are shown for the two treatment arms in Figure~\ref{fig:DFS}.

\begin{figure}[t!]
\begin{Schunk}

{\centering \includegraphics{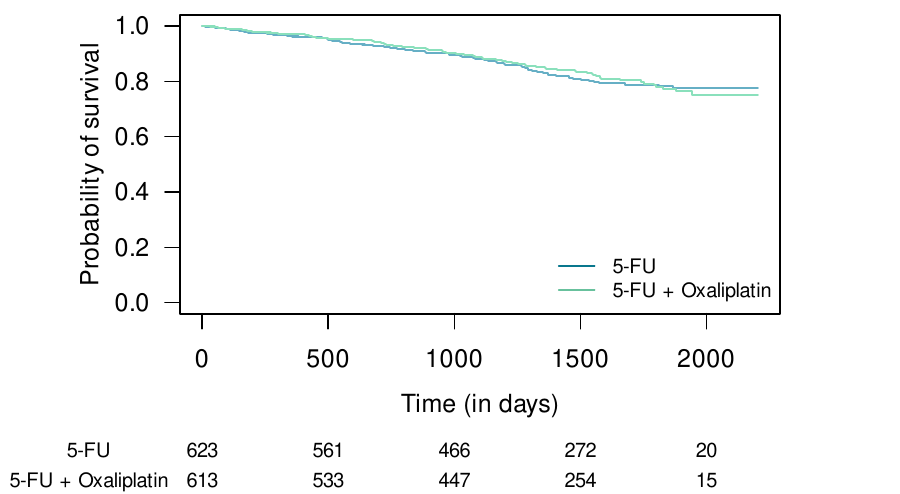} 

}

\end{Schunk}
\caption{Overall survival. The survivor functions of the two treatment arms estimated
by the non-parametric Kaplan-Meier method are shown together with the number at risk table.}
\label{fig:OS}
\end{figure}

A secondary endpoint considered in the study is overall survival, defined as
the time $\rY \in \RR^{+}$ from randomisation to death
from any cause.  Notably, all observations $\ry$ for this endpoint are
exact or right-censored.  The corresponding survivor curves, estimated
non-parametrically by the Kaplan-Meier method \citep{Kaplan_Meier_1958}, are shown in Figure~\ref{fig:OS}.

The primary data analysis for this trial was performed by
\citet{Roedel_Graeven_Fietkau_2015}.  In their analysis, the
treatment effect comparing the effect of the experimental treatment
to the effect of the control treatment on disease-free and overall survival was
assessed by adjusted log-rank tests and mixed-effects Cox models (both adjusting for the stratified randomisation process), treating both survival endpoints as exact.  Following, we demonstrate
the process of fitting a fully parametric mixed-effects Cox model that
accounts for interval-censored event times of the primary endpoint in Section~\ref{sec:dObs}, also in terms
of a model featuring marginal interpretation of the estimated hazard ratio.

A subsequent post hoc analysis was carried out by
\citet{Hofheinz_Arnold_Fokas_2018}.  In this analysis, the
possibility of an age-varying treatment effect on both the primary endpoint
of disease-free survival and the secondary endpoint of overall survival was
investigated.  In
Section~\ref{sec:PM}, we demonstrate how such an analysis can be performed
within the discussed framework, taking into account interval-censoring.
We illustrate two approaches for estimating age-varying effects, using smooth
age-varying hazard ratios and age-structured survival trees.

While the analyses conducted by \citet{Roedel_Graeven_Fietkau_2015} and \citet{Hofheinz_Arnold_Fokas_2018}
made serious efforts to address the research
questions effectively, they were limited by the lack of software resources capable
of adequately handling interval-censored and correlated observations for the analysis of the primary endpoint.
Notably,
the first \proglang{R} add-on package capable of estimating Cox models in
the presence of interval-censoring was published in 2014
(\citealp[\pkg{coxinterval} package][]{pkg:coxinterval}). At the time of the
statistical analysis of the primary endpoint, it was impossible to fit
mixed-effects models with flexible baseline hazards to interval-censored outcomes.
This obstacle was one of the main
motivation to develop a comprehensive software package implementing a
general class of transformation models with applications in the domain of survival
analysis. The corresponding framework implementing smooth
transformation models \citep{Hothorn_Moest_Buehlmann_2017} in \proglang{R}
helps to address such and other practically relevant limitations.
In this tutorial, we present analyses that the CAO/ARO/AIO-04 study investigators
would have liked to have been able to perform a decade ago.

\section{Independent observations} \label{sec:iObs}

\subsection{Survival models}
\subsubsection{Weibull proportional hazards model} \label{sec:WEI}

Probably one of the most renowned parametric model in survival analysis
is the Weibull model
\citep{Wei_1992}, where the response $\rY$ conditional on treatment
assignment $\rW = \rw$ is assumed to follow a Weibull
distribution.  We consider the Weibull proportional hazards model with
survivor functions conditional on the treatment assignment parametrised as,
\begin{eqnarray*}
\snul(\ry) &=& \Prob(\rY > \ry \mid \rW = 0) =
\sMEV{\eparm_1 + \eparm_2 \log(\ry)},
\qquad \qquad \eparm_2 > 0,\nonumber\\
\sone(\ry) &=& \Prob(\rY > \ry \mid \rW = 1) =
\sMEV{\eparm_1 + \eparm_2 \log(\ry) - \eshiftparm},\nonumber\\
\end{eqnarray*}
with the general formula,
\begin{eqnarray} \label{mod:WEI}
\sw(\ry) &=& \Prob(\rY > \ry \mid \rW = \rw) =
\sMEV{\eparm_1 + \eparm_2 \log(\ry) - \eshiftparm \rw}.
\end{eqnarray}
The log-cumulative baseline hazard $\log(-\log(\snul(\ry))) =
\log(\Haznul(\ry))$ here is given by $\h(\ry) = \eparm_1 + \eparm_2
\log(\ry)$, assuming a linear shift $\eshiftparm$ on the scale of log-time $\log(\ry)$.  The model not only assumes proportional
hazards, with hazard ratio $\nicefrac{\Hazone(\ry)}{\Haznul(\ry)} =
\nicefrac{\hazone(\ry)}{\haznul(\ry)} = \exp(-\eshiftparm)$, but is also an
accelerated failure time model
\begin{eqnarray*} \label{mod:WEIAFT}
\log(\rY) &=& \frac{- \eparm_1 + \eshiftparm \rw + \rZ}{\eparm_2}, \qquad \rZ \sim \MEV\\
&=& - \frac{\eparm_1}{\eparm_2} + \frac{\eshiftparm}{\eparm_2} \rw + \frac{1}{\eparm_2}\rZ
= \alpha + \tilde\eshiftparm \rw + \sigma \rZ,
\end{eqnarray*}
with the errors $\rZ$ following a minimum extreme value distribution (MEV). 
Consequently, $\rY \mid \rW = \rw$ follows a Weibull distribution
(\citealt{Kalbfleisch_Prentice_2002}, Chapter~2) with intercept
$\alpha = -\eparm_1 \eparm_2^{-1}$, scale parameter $\sigma = \eparm_2^{-1}$ and log-acceleration
factor $\tilde\eshiftparm = \eshiftparm \eparm_2^{-1}$. 
The model implies that,
for the experimental arm time $\rY$ is accelerated by
$\exp(\tilde\eshiftparm)$, that is $\rY_1 = \exp(\tilde\eshiftparm) \rY_0$,
thus the probability of disease-free survival for the experimental arm is given by
$\sone(\ry) = \snul(\exp(-\tilde\eshiftparm) \ry)$.

Alternatively, different distributions for $\rZ$, such as the normal or
logistic distribution, can be specified, leading to the formulation of
log-normal or log-logistic models.

Parameter estimation of the Weibull model is straightforward using maximum
likelihood, because the distribution function can be directly evaluated and
thus allows to effectively handle interval-censored or truncated data, as
will be discussed in Section~\ref{subsec:ll}.

\subsubsection{Flexible proportional hazards model} \label{sec:COX}

The assumption of a log-linear log-cumulative baseline hazard function $\h$, implied by the
Weibull model, can be relaxed by replacing
$\log(\Haznul(\ry)) = \h(\ry) = \eparm_1 + \eparm_2
\log(\ry)$ with a more flexible function
$\h(\ry) = \basisy(\ry)^\top \parm$ defined in terms of smooth spline basis
functions $\basisy$ and corresponding parameters $\parm$.  This yields the following model
\begin{eqnarray} \label{mod:COX}
\sw(\ry) &=& \Prob(\rY > \ry \mid \rW = \rw) = \sMEV{\h(\ry) + \eshiftparm \rw},
\end{eqnarray}
where the hazard ratio is given by $\nicefrac{\Hazone(\ry)}{\Haznul(\ry)} =
\nicefrac{\hazone(\ry)}{\haznul(\ry)} = \exp(\eshiftparm)$. 
\citet{McLain_Ghosh_2013} and \citet{Hothorn_Moest_Buehlmann_2017} proposed
parametrising $\h(\ry) = \basisy(\ry)^\top \parm$ as polynomials in
Bernstein form $\h(\ry) = \bern{\dimparm - 1}(\ry)^\top \parm$ of order
$\dimparm - 1$.\deleted{In the context of skewed survival data, it might be
suitable to apply polynomials in Bernstein form to the logarithm of the survival
time $\ry$, leading to $\h(\ry) = \bern{\dimparm - 1}(\log(\ry))^\top
\parm$.} The corresponding model~(\ref{mod:COX}) is a fully parametric version of the otherwise semi-parametric Cox
proportional hazards model \citep{Cox_1972}.  The latter treats $\h$ as an
infinite dimensional nuisance parameter which is profiled out from the
likelihood.  This leads to the partial likelihood,
through which the log-hazard ratio $\eshiftparm$ can be estimated \citep{Cox_1975}.  In contrast, all
parameters of model (\ref{mod:COX}) are estimated simultaneously by maximum
likelihood.  The parametrisation of $\h$ in terms of basis functions
and corresponding parameters allows to specify of a flexible, yet fully
parametric, monotonically increasing log-cumulative hazard function. 
This is achieved under appropriate constraints $\eparm_p \le \eparm_{p + 1}$
for $p \in 1, \dots, \dimparm - 1$ \citep{Hothorn_Moest_Buehlmann_2017}. 
Adopting this specific parametrisation for the log-cumulative baseline
hazard function $\log(\Haznul(\ry)) = \h(\ry)$ facilitates the computation
of the corresponding density $\dnul(\ry)$ and distribution function
$\pnul(\ry)$, thus allowing for straightforward parameter estimation using
maximum likelihood.  This holds true even when dealing with
interval-censored or truncated observations.

Within the \pkg{tram} add-on package, the order, $\dimparm - 1$, of
polynomials in Bernstein form is not determined in a data-driven way. The default $\dimparm -
1 = 6$ is typically a good compromise between flexibility of $\h(\ry)$ and
computing time needed to optimise the log-likelihood. Fixed $\dimparm$ also facilitates
standard maximum likelihood inference. Because of the
monotonicity constraint on $\h$, the transformation function 
$\h$ it not prone to overfit and
thus, in principle, $\dimparm$ can be chosen much larger.
The effect of larger $\dimparm$ on 
estimates of other model parameters and their standard errors is very small
(see, for example, the log-hazard ratios in Figure~5 provided by \citet{Hothorn_2018}
and the empirical comparison to non-parametric models
\cite{Yuqi_Hothorn_Li_2020}). However, if one is
interested in expressions involving the derivative $\h^\prime(\ry)$, which
itself is in Bernstein form, the order $\dimparm - 1$ must be chosen
in a data-driven way, for example for density estimation. Sieve maximum
likelihood procedures have been suggested in this context, for example in Cox models with
log-cumulative baseline hazard functions in Bernstein form
\cite{McLain_Ghosh_2013}.

\subsubsection{Stratified proportional hazards model} \label{sec:STRAT}

Accounting for variations in baseline risks among different patient
strata identified by variable $\rs$, one can employ stratified models that
incorporate stratum-specific log-cumulative baseline hazard functions
$\h(\ry \mid \rs)$.  These models can be defined by
\begin{eqnarray*} \label{mod:STRAT}
\sw(\ry \mid \rs) &=& \Prob(\rY > \ry \mid \rS = \rs, \rW = \rw) =
\sMEV{\h(\ry \mid \rs) + \eshiftparm \rw},
\end{eqnarray*}
with $\h(\ry \mid \rs) = \basisy(\ry)^\top \parm(\rs)$ and global hazard
ratio $\nicefrac{\Hazone(\ry \mid \rs)}{\Haznul(\ry \mid \rs)} =
\nicefrac{\hazone(\ry \mid \rs)}{\haznul(\ry \mid \rs)} = \exp(\eshiftparm)$,
assuming that the treatment effect is the same
across all patient strata $\rs$.

One could further relax the restriction of a global treatment effect, allowing
for an interaction of the treatment assignment $\rW = \rw$ and the stratum
$\rs$ by formulating the log-cumulative hazard as $\h(\ry \mid \rs) + \rw
\eshiftparm + \gammavec^\top(\rw \times \rs)$.

\subsubsection{Non-proportional hazards model} \label{sec:SCOX}

Extending beyond the proportional hazards assumption, an additional
treatment-dependent model term can be estimated.  \citet{Burke_2017}
introduced the multi-parameter extension to the Weibull model~(\ref{mod:WEI}) in the context
of survival analysis, specifically outlining its use for interval-censored observations \citep{Burke_SiM_2020}.

A similar extension can be made to the more flexible, parametric, Cox
model~(\ref{mod:COX}),
by additionally estimating a scale term $\escaleparm$ for the experimental arm
\citep{Siegfried_Kook_Hothorn_2023},
\begin{eqnarray*} \label{mod:SCOX}
\sw(\ry) &=& \Prob(\rY > \ry \mid \rW = \rw) =
\sMEV{\sqrt{\exp(\escaleparm\rw)}\,\h(\ry) + \eshiftparm \rw}.
\end{eqnarray*}
In this case the ratio of the cumulative hazards, $\nicefrac{\Hazone(\ry)}{\Haznul(\ry)}$, is a
non-proportional function of $\ry$. 
The corresponding bivariate score test 
(Section~3.1.2 of \citealt{Siegfried_Kook_Hothorn_2023}) further allows to test the null 
hypothesis of equal survival, \ie~$\eshiftparm = \escaleparm = 0$, 
without relying on the assumption of proportional hazards.

\subsubsection{Time-varying hazards model} \label{sec:TCOX}

Accounting for changing effects of the treatment over time, we can further
extend beyond the proportional hazards assumption and estimate a model
incorporating a time-varying treatment effect,
\begin{eqnarray*} \label{mod:TCOX}
\sw(\ry) &=& \Prob(\rY > \ry \mid \rW = \rw) =
\sMEV{\h(\ry) + \eshiftparm(\ry)\rw}.
\end{eqnarray*}
Here, the model introduces a time-varying shift $\eshiftparm(\ry)$ in the log-cumulative hazard
function $\log(\Hazone(\ry)) = \log(\Haznul(\ry)) + \eshiftparm(\ry)$, thereby relaxing
the assumption of a constant log-hazard ratio $\eshiftparm$. The shift
$\eshiftparm(\ry)$ is also parameterised in terms of a
polynomial in Bernstein form, thus allowing to estimate a time-varying ratio
of the cumulative hazards $\nicefrac{\Hazone(\ry)}{\Haznul(\ry)} = \exp(\eshiftparm(\ry))$.

\subsection{Likelihood} \label{subsec:ll}

From the above models, the log-likelihoods for exact or independently right-,
left- or interval-censored and truncated observations are easily deducible. 
We here consider the most general case where the log-cumulative hazard
function is given by $\h(\ry \mid \rw, \rs) = \sqrt{\exp(\escaleparm \rw)}\,\h(\ry
\mid \rs) + \eshiftparm \rw$.

For exact continuous observations $(\ry, \rw, \rs)$, the corresponding likelihood
contributions are given by
\begin{eqnarray*} \label{eq:cll}
\ell(\parm(\rs), \eshiftparm, \escaleparm \mid \rY = \ry) =
  \log\left\{\dZ\left[\h(\ry \mid \rw, \rs) \right]\right\} +
  \log\left\{\h^\prime(\ry \mid \rw, \rs)\right\},
\end{eqnarray*}
with the standard minimum extreme value density $\dZ(\rz) = \exp(\rz -
\exp(\rz))$ and the derivative of the log-cumulative hazard function with
respect to $\ry$, $\h^\prime(\ry \mid \rw, \rs) = \sqrt{\exp(
\escaleparm \rw)}\, \basisy^\prime(\ry)^\top \parm(\rs)$.

Because the transformation function $\h$, defining the log-cumulative
baseline hazard function, is parametrised in terms of polynomials in
Bernstein form, where the basis functions $\bern{\dimparm - 1}(\ry) \in
\RR^\dimparm$ are specific beta densities \citep{Farouki_2012}, it is
straightforward to obtain the derivatives of the basis functions with
respect to $\ry$, leading to $\h^\prime(\ry \mid \rs) = 
\bern{\dimparm - 1}^\prime(\ry)^\top \parm(\rs)$.

Under independent left-, right- or interval-censored event times
$(\ubar{\ry}, \bar{\ry}]$ the exact log-likelihood contribution is
\begin{eqnarray*} \label{eq:ill}
\ell(\parm(\rs), \eshiftparm, \escaleparm \mid \rY \in (\ubar{\ry}, \bar{\ry}])  =
\log\left\{\Prob(\rY \in (\ubar{\ry}, \bar{\ry}] \mid \rw, \rs)\right\} = 
\log\left\{\sw(\ubar\ry \mid \rw, \rs) -
     \sw(\bar\ry \mid \rw, \rs)\right\}.
\end{eqnarray*}
For observations that are right-censored at time $\ry$ the datum is given by
$(\ubar{\ry}, \bar{\ry}] = (\ry, \infty)$ and for left-censored observations
it is $(\ubar{\ry}, \bar{\ry}] = (0, \ry]$.

In cases where event times are subject to random left-, right-, or
interval-truncation $(\ry_{\text{l}}, \ry_{\text{r}}] \subset \RR^{+}$, the
above log-likelihood contributions change to
\begin{eqnarray*} \label{eq:tll}
\ell(\parm(\rs), \eshiftparm, \escaleparm \mid \rY \in (\ubar{\ry}, \bar{\ry}]) -
\ell(\parm(\rs), \eshiftparm, \escaleparm \mid \rY \in (\ry_\text{l}, \ry_\text{r}])
\end{eqnarray*}
with $\ry_\text{l} = 0$ for right-truncated and $\ry_\text{r} = \infty$ for
left-truncated observations.  Such considerations are relevant in scenarios
involving a late entry approach, for instance, resulting in left-truncated
observations, where one is interested in modelling $\Prob(\ry > \rY \mid \rY \in
(\ry_{\text{l}}, \infty))$, or for modelling time-varying covariates.

%
\subsection{Application}

Now, turning our attention to the CAO/ARO/AIO-04 two-arm
randomised controlled trial, we aim to estimate the previously discussed
models using the unified maximum likelihood framework provided by the
\proglang{R} add-on package \pkg{tram} \citep{pkg:tram}.  We fit the models
to the primary endpoint of disease-free survival $\rY$ estimating the treatment
effect corresponding to the assignment $\rW =
\var{randarm}$. 

The disease-free survival times are stored as \code{iDFS},
an object of class `\code{Surv}',
which includes a mix of exact, interval-, and right-censored observations.
This `\code{Surv}' object can be specified with
\code{Surv(CAOsurv\$iDFStime, CAOsurv\$iDFStime2, type = "interval2")}
\citep{pkg:survival,Therneau_Grambsch_2000}.
Exact observations are represented by two identical time points,
for interval-censored observations, the two times define the period within
which the event occurred and right-censored observations are represented by an
interval from the last visit to infinity.

We start by fitting the Weibull model (Section~\ref{sec:WEI}) using the \cmd{Survreg}
function.
\begin{Schunk}
\begin{Sinput}
R> Survreg(iDFS ~ randarm, data = CAOsurv, dist = "weibull")
\end{Sinput}
\end{Schunk}

\begin{center}

\scalebox{0.8}{
\begin{tabular}{rrrr}
  \toprule
Coefficient & Estimate & Std. Error & 95\%-Wald CI \\ 
  \midrule
$\eparm_1$ & $ -6.231 $ & 0.265 & $-6.752$ to $-5.711$ \\ 
  $\eparm_2$ & $ 0.733 $ & 0.036 & $\phantom{-}0.663$ to $\phantom{-}0.803$ \\ 
  $\eshiftparm$ & $ 0.229 $ & 0.106 & $\phantom{-}0.020$ to $\phantom{-}0.438$ \\ 
   \\[-1.2ex] \toprule Log-Likelihood & Likelihood Ratio Test & Score Test & Permutation Score Test \\ 
   \midrule $-$2'281.17 & $p$ = 0.031 & $p$ = 0.031 & $p$ = 0.035 \\ 
   \\[-2ex]  \bottomrule
\end{tabular}
}

\end{center}

The model quantifies the treatment effect through a hazard ratio
$\exp(-\hat\eshiftparm) = 0.795$, comparing the hazards of the
experimental arm to the hazards of the control arm.  The results indicate
a reduction in hazards for patients receiving the
experimental treatment compared to the control treatment. 
This reduction in hazards translates to a
prolonged disease-free survival time in the experimental arm. 
Since the model is a proportional hazards counterpart of the Weibull accelerated
failure time model fitted by \cmd{survreg} from the \pkg{survival} package
\citep{Therneau_Grambsch_2000,pkg:survival},
the estimate can also be translated into a log-acceleration factor
$\hat{\tilde\eshiftparm} = \hat\eshiftparm \hat\eparm_2^{-1} =
0.312$.  This implies that
the disease-free survival time $\rY$ is prolonged by
$\exp(\hat{\tilde\eshiftparm}) = 1.367$ in the
experimental arm, compared to the control arm. 

Next, we fit the flexible proportional hazards model (Section~\ref{sec:COX}) using the
\cmd{Coxph} function from the \pkg{tram} package.
\begin{Schunk}
\begin{Sinput}
R> Coxph(iDFS ~ randarm, data = CAOsurv)
\end{Sinput}
\end{Schunk}

\begin{center}

\scalebox{0.8}{
\begin{tabular}{rrrr}
  \toprule
Coefficient & Estimate & Std. Error & 95\%-Score CI \\ 
  \midrule
$\eshiftparm$ & $ -0.231 $ & 0.107 & $-0.439$ to $-0.022$ \\ 
   \\[-1.2ex] \toprule Log-Likelihood & Likelihood Ratio Test & Score Test & Permutation Score Test \\ 
   \midrule $-$2'242.25 & $p$ = 0.030 & $p$ = 0.030 & $p$ = 0.030 \\ 
   \\[-2ex]  \bottomrule
\end{tabular}
}

\end{center}

\begin{figure}[t!]
\centering

\begin{Schunk}

{\centering \includegraphics{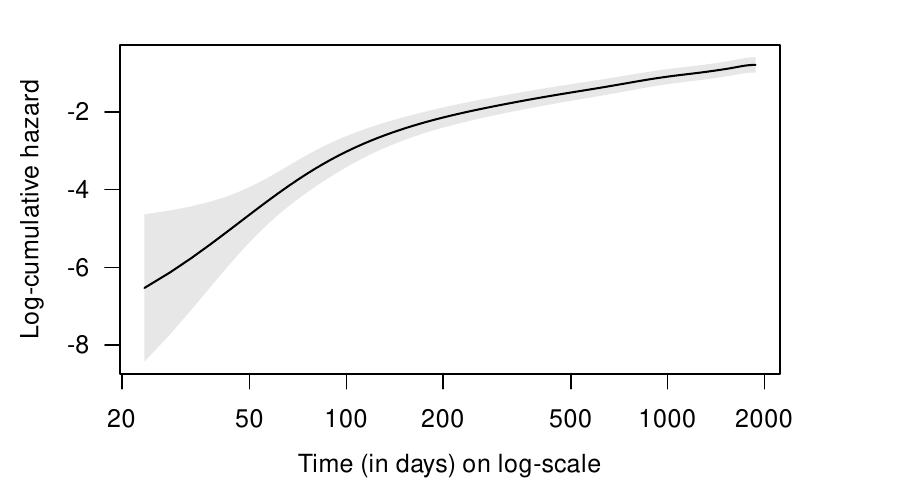} 

}

\end{Schunk}
\caption{Flexible proportional hazards model.
Log-cumulative baseline hazard function and corresponding 95\%-confidence
band estimated by the model.} \label{fig:COX}
\end{figure}

The fitted model is a fully parametric version of the famous Cox model,
otherwise estimated semi-parametrically using the partial likelihood (as
implemented in the \pkg{survival} package in the \cmd{coxph} function). 
Here the log-cumulative hazard function is specified in terms of
polynomials in Bernstein form, by default of order $\dimparm - 1 = 6$, 
\deleted{which is defined on the scale of $\log(\ry)$ (setting \code{log\_first =
TRUE}),}
specifying the transformation function $\h(\ry) = \bern{6}(\ry)^\top\parm$. 
The fully parametric approach enables the
straightforward incorporation of interval-censored disease-free survival times. 
Figure~\ref{fig:COX} illustrates the estimated log-cumulative baseline hazard
function $\log(\hatHaznul(\ry)) = \hat{\h}(\ry) =
\bern{6}(\ry)^\top\hat{\parm}$ along with the $95\%$-confidence band,
revealing a non-linear function of log-time. The band was
obtained from simultaneous confidence intervals for a dense grid of time
points utilising the asymptotic normality of the maximum likelihood
estimator $\hat{\parm}$ and the fact that $\h$ was parameterised as
a contrast.
The estimated hazard ratio is $\exp(\hat\eshiftparm) = 0.794$,
indicating reduced hazards in the experimental arm.
The software implementation further allows
to compute a corresponding 95\%-permutation score confidence interval which
ranges from $0.645$ to $0.978$.

To further accommodate for varying log-cumulative baseline hazard functions $\Haznul(\ry \mid \rs)$
across patient strata $\rs$ (here identified by $\code{strat}$), we can fit
a stratified model (Section~\ref{sec:STRAT}).
\begin{Schunk}
\begin{Sinput}
R> Coxph(iDFS | strat ~ randarm, data = CAOsurv)
\end{Sinput}
\end{Schunk}

\begin{center}

\scalebox{0.8}{
\begin{tabular}{rrrr}
  \toprule
Coefficient & Estimate & Std. Error & 95\%-Score CI \\ 
  \midrule
$\eshiftparm$ & $ -0.228 $ & 0.107 & $-0.436$ to $-0.019$ \\ 
   \\[-1.2ex] \toprule Log-Likelihood & Likelihood Ratio Test & Score Test & Permutation Score Test \\ 
   \midrule $-$2'213.94 & $p$ = 0.031 & $p$ = 0.032 & $p$ = 0.033 \\ 
   \\[-2ex]  \bottomrule
\end{tabular}
}

\end{center}

\begin{figure}[t!]
\centering
\begin{Schunk}

{\centering \includegraphics{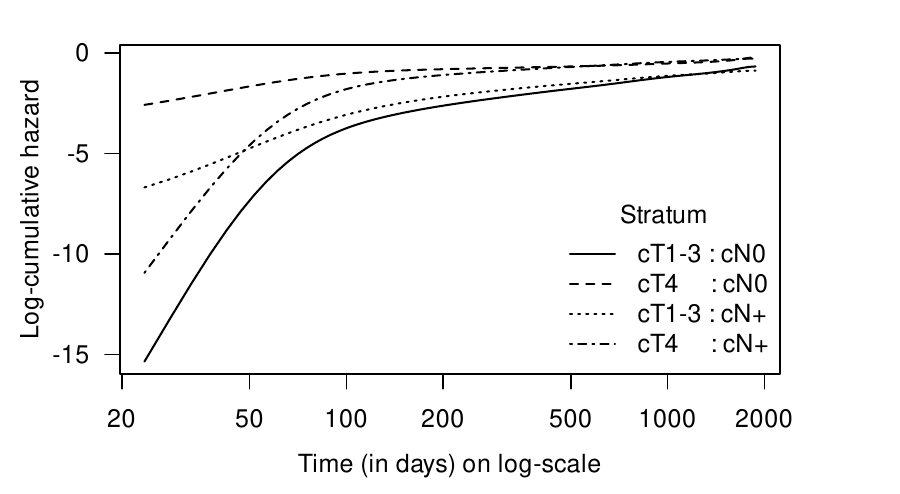} 

}

\end{Schunk}

\caption{Stratified proportional hazards model.
Log-cumulative baseline hazard functions $\Haznul(\ry \mid \rs)$ estimated
by the model are shown separately for each stratum $\rs$.}\label{fig:STRAT}
\end{figure}

The model estimates separate smooth log-cumulative baseline hazard functions for each
stratum~$\rs$, as illustrated in Figure~\ref{fig:STRAT}, but provides an estimate
of the global hazard ratio $\exp(\hat\eshiftparm) = 0.796$,
indicating a reduction of the hazard in the experimental arm by
0.796 relative to the hazard in the control arm across all
stratum.

Moving away from the proportional hazards assumption, we can fit a non-proportional hazards model
(a location-scale version of the Cox model, Section~\ref{sec:SCOX}) using the same function.
\begin{Schunk}
\begin{Sinput}
R> Coxph(iDFS ~ randarm | randarm, data = CAOsurv)
\end{Sinput}
\end{Schunk}

\begin{center}

\scalebox{0.8}{
\begin{tabular}{rrrr}
  \toprule
Coefficient & Estimate & Std. Error & 95\%-Wald CI \\ 
  \midrule
$\eshiftparm$ & $ -0.091 $ & 0.163 & $-0.411$ to $0.229$ \\ 
  $\escaleparm$ & $ 0.257 $ & 0.203 & $-0.140$ to $0.654$ \\ 
   \\[-1.2ex] \toprule Log-Likelihood & Likelihood Ratio Test & Bivariate Permutation Score Test &   \\ 
   \midrule $-$2'241.46 & $p$ = 0.043 & $p$ = 0.027 &  \\ 
   \\[-2ex]  \bottomrule
\end{tabular}
}

\end{center}

\begin{figure}[t!]
\centering
\begin{Schunk}

{\centering \includegraphics{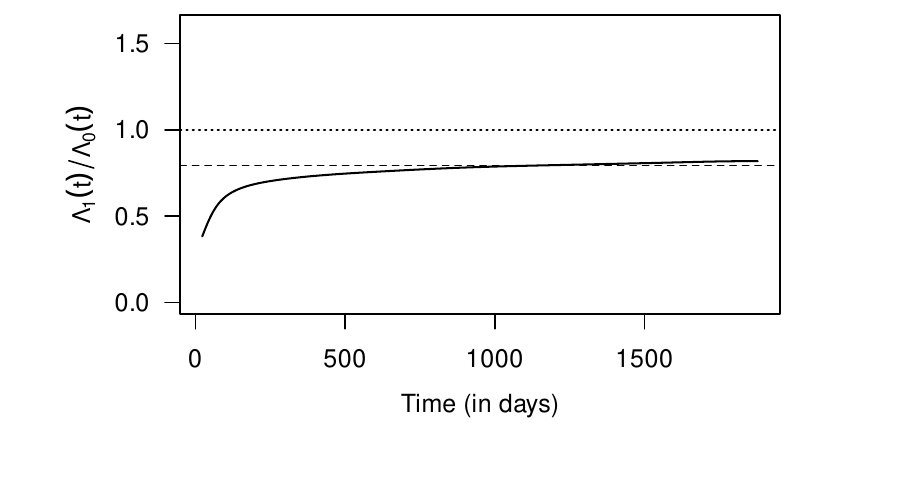} 

}

\end{Schunk}

\caption{Non-proportional hazards model. $\nicefrac{\Hazone(\ry)}{\Haznul(\ry)}$ (solid line) estimated by the model is shown alongside the constant hazard
ratio estimated from the proportional hazards model (dashed line) over time
$\ry$.}\label{fig:SCOX}
\end{figure}

The ratio of the cumulative hazards $\nicefrac{\widehat\Hazone(\ry)}{\widehat\Haznul(\ry)}$,
shown in Figure~\ref{fig:SCOX}, no longer remains proportional but varies over time. 
The curve indicates a pronounced initial reduction in cumulative hazards for the
experimental arm compared to the control arm, which
gradually decreases as time progresses.  This suggests that the treatment
effect is stronger in the beginning.  The corresponding bivariate score
test, which tests the null hypothesis of equal survival without assuming
proportional hazards, further indicates evidence for non-equal disease-free
survival times.

Finally, we fit the model featuring a time-varying treatment effect (Section~\ref{sec:TCOX}).
\begin{Schunk}
\begin{Sinput}
R> Coxph(iDFS | randarm ~ 1, data = CAOsurv)
\end{Sinput}
\end{Schunk}

\begin{figure}[t!]
\centering

\begin{Schunk}

{\centering \includegraphics{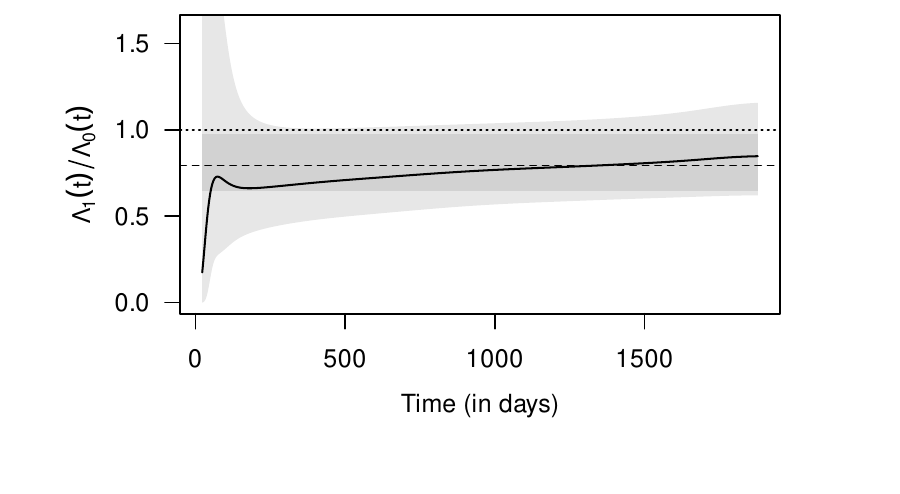} 

}

\end{Schunk}
\caption{Time-varying hazards model. $\nicefrac{\Hazone(\ry)}{\Haznul(\ry)}$
and corresponding 95\%-confidence
bands over time
$\ry$ (solid line) estimated by the model is shown alongside the constant hazard
ratio estimated from the proportional hazards model (dashed line). The log-likelihood of the model is
$-$2'240.21.} \label{fig:TCOX}
\end{figure}

The treatment effect $\nicefrac{\Hazone(\ry)}{\Haznul(\ry)} =
\exp(\eshiftparm(\ry))$ is a function of time, as shown in
Figure~\ref{fig:TCOX}.  The curve again demonstrates a reduction in hazards
for the experimental arm compared to the control arm, which is more
substantial in the beginning and gradually becomes less prominent as time
progresses.

\newpage

\section{Dependent censoring} \label{sec:dCens}

\subsection{Copula proportional hazards model}

Until now, the models, we have discussed, have been constructed under the
assumption of independent censoring, which implies that the censoring times
$\rC$ are independent of the event times $\rY$ given the treatment
assignment $\rW = \rw$, that is $\rY \indep \rC \mid \rW = \rw$.  We can
however move beyond relying on this assumption and allow the model to
capture potential dependence between the censoring times $\rC$ and event times $\rY$.

We explore models discussed in recent work of \citet{Cazado_Keilegom_2022},
which involve a joint model for $\rY$ and $\rC$ employing a bivariate
Gaussian copula of 
the marginal survivor functions $\sYw(\ry)$ and 
$\sCw(\rc)$ of $\rY$ and $\rC$ respectively,
\begin{eqnarray*} \label{mod:DEPC}
\Prob(\rY \leq \ry, \rC \leq \rc \mid \rW = \rw) = \Phi_{0, \mR(\lparm)}
\left\{
\Phi^{-1}\left[1 - \sYw(\ry)\right],
\Phi^{-1}\left[1 - \sCw(\rc)\right]
\right\}
\end{eqnarray*}
with correlation matrix
\begin{eqnarray*} \label{mod:DEPC:CORR}
\mR(\lparm) =
\left[\begin{array}{cc}
1 & \nicefrac{-\lparm}{\sqrt{1 + \lparm^2}}\\
\nicefrac{-\lparm}{\sqrt{1 + \lparm^2}} & 1
\end{array} \right], \qquad \lparm \in (-\infty, \infty)
\end{eqnarray*}
to account for the association between $\rY$ and $\rC$.
\citet{Deresa_Keilegom_2023} recently demonstrated that the above model
maintains identifiability when the marginal survivor functions $\sYw(\ry)$ and
$\sCw(\rc)$ are described by a
flexible proportional hazards model~(\ref{mod:COX}) and a model that assumes
a Weibull distribution~(\ref{mod:WEI}),
respectively.  This allows to estimate the dependence parameter $\lparm$ and
other model parameters from the observed data.
A dependence parameter $\lparm$ of zero corresponds to uncorrelated event
times $\rY$ and censoring times $\rC$, thus indicating lack of evidence for dependent censoring.

\subsection{Application}

Returning to our application, where we previously assumed independent
censoring of disease-free survival times,
we now aim to address the potential scenario where loss of
follow-up time $\rC \in \RR^{+}$ in the two arms is not independent of the
overall survival time  $\rY \in \RR^{+}$ (secondary endpoint).

The observed times can be categorised into the following event types (specified in \code{DepCevent}):
The event of interest (corresponding to overall survival), loss of
follow-up (dependent censoring), and end of follow-up period
(administrative/independent censoring).
\begin{center}
\small

\begin{tabular}{lrr}
  \toprule
 & 5-FU & 5-FU + Oxaliplatin \\ 
  \midrule
Administrative censoring & 469 & 466 \\ 
  Event of interest & 106 &  96 \\ 
  Loss of follow-up &  48 &  51 \\ 
   \bottomrule
\end{tabular}

\end{center}
The model accommodating dependent censoring can also be fitted using the
\cmd{Coxph} function by appropriately specifying the event in the
`\code{Surv}' object.
\begin{Schunk}
\begin{Sinput}
R> Coxph(Surv(OStime, event = DepCevent) ~ randarm, data = CAOsurv)
\end{Sinput}
\end{Schunk}

\begin{center}

\scalebox{0.8}{
\begin{tabular}{rrrr}
  \toprule
Coefficient & Estimate & Std. Error & 95\%-Wald CI \\ 
  \midrule
$\eshiftparm_\rY$ & $ -0.031 $ & 0.143 & $-0.311$ to $0.248$ \\ 
  $\lparm$ & $ 0.021 $ & 0.452 & $-0.865$ to $0.908$ \\ 
   \\[-1.2ex] \toprule Log-Likelihood &   &   &   \\ 
   \midrule $-$3'061.65 &  &  &  \\ 
   \\[-2ex]  \bottomrule
\end{tabular}
}

\end{center}

The joint model is fitted based on a Gaussian copula, estimating a marginal
flexible proportional hazards model~(\ref{mod:COX}) for the overall survival time $\rY$
and a marginal Weibull proportional hazards model~(\ref{mod:WEI})
for the loss of follow-up time $\rC$, while
accounting for independent right-censoring at the end of the follow-up
period.

The estimated hazard ratio assessing the treatment effect on overall survival,
is $\exp(\hat\eshiftparm_\rY) = 0.969$ with a
95\%-confidence interval from $0.733$ to $1.282$.  This indicates
no evidence for prolonged overall survival in the experimental compared to the control arm.  The estimated dependence
parameter is $\hat\lparm = 0.021$,
corresponding to a Kendall's $\hat\tau = -0.014$. 
The corresponding 95\%-confidence interval from $-0.865$ to $0.908$
for $\lparm$ does include zero, providing no evidence for a dependence between time
of overall survival $\rY$ and loss of follow-up $\rC$ given the treatment assignment $\rW = \rw$.

\section{Dependent observations} \label{sec:dObs}

\subsection{Survival models}

Up to this point, the models we have discussed have been built upon the
assumption of independent observations.  However, this
assumption may not hold in situations where observations are clustered, such
as for multi-center trials where observations from the same center are generally
correlated.

\subsubsection{Mixed-effects proportional hazards model} \label{sec:COXME}

In order to address this challenge, we can leverage a flexible mixed-effects
proportional hazards model as proposed by \citet{Tamasi_Crowther_Puhan_2022}. 
This approach extends the previously discussed smooth transformation models by conditioning on an
unobserved cluster-specific random effect $R = r$ that accounts for the dependence within
clusters,
\begin{eqnarray*} \label{mod:COXME}
\sw(\ry \mid R = r) &=&  \Prob(\rY > \ry \mid \rW = \rw, R = r) = 
\sMEV{\h(\ry) + \eshiftparm \rw + r}.
\end{eqnarray*}
This formulation provides a fully parametric version of the Cox proportional
hazards model~(\ref{mod:COX}), incorporating multivariate normally distributed random
effects with a zero mean and variance $\tau^2$.  The treatment
effect $\eshiftparm$ is interpreted as a log-hazard ratio conditional on
unobserved random effects. For more in-depth information on likelihood-based inference, 
see \citet{Tamasi_Hothorn_2021} and \citet{Tamasi_Crowther_Puhan_2022}.

\subsubsection{Marginalised proportional hazards model} \label{sec:MCOX}

Furthermore, we can explore the model proposed by
\citet{Barbanti_Hothorn_2023}, where the marginal survivor functions are
characterised by models~(\ref{mod:COX}), while the correlations within
clusters are captured by a joint multivariate normal distribution.  This
joint modeling approach yields a marginalised formulation for the random
intercept model,
\begin{eqnarray*} \label{mod:MCOX}
\sw(\ry) &=&  \Prob(\rY > \ry \mid \rW = \rw) =
\sMEV{\frac{\h(\ry) + \eshiftparm \rw}{\sqrt{\mparm^2 + 1}}}.\\
\end{eqnarray*}
Here, $\mparm^2$ represents the variance of a cluster-specific intercept. 
Using this framework, it becomes possible to quantify the treatment effect
using the marginal hazard ratio given by $\exp\left(\nicefrac{\eshiftparm}{
\sqrt{\mparm^2 + 1}}\right)$.

Further details on the models, including likelihood-based inference, can be
found in \citet{Barbanti_Hothorn_2023}.

\subsection{Application}

To estimate mixed-effects smooth transformation models (Section~\ref{sec:COXME}) we can use the
\pkg{tramME} package \citep{pkg:tramME,Tamasi_Hothorn_2021}, available from 
CRAN:
\begin{Schunk}
\begin{Sinput}
R> install.packages("tramME")
R> library("tramME")
\end{Sinput}
\end{Schunk}
Including a random-intercept for the block used in the randomisation, which
is a combination of the centers $j = 1, \dots
88$ and the stratum $\rs$ ($j \times
\rs = \code{Block}$) in the model, we can account for potential correlation
between patients from the same block.  The corresponding mixed-effects
proportional hazards model can be estimated using the \code{CoxphME()}
function.
\begin{Schunk}
\begin{Sinput}
R> CoxphME(iDFS ~ randarm + (1 | Block), data = CAOsurv)
\end{Sinput}
\end{Schunk}

\begin{center}

\scalebox{0.8}{
\begin{tabular}{rrrr}
  \toprule
Coefficient & Estimate & Std. Error & 95\%-Wald CI \\ 
  \midrule
$\eshiftparm$ & $ -0.235 $ & 0.107 & $-0.445$ to $-0.026$ \\ 
  $\tau^2$ & 0.071 &  &  \\ 
   \\[-1.2ex] \toprule Log-Likelihood &   &   &   \\ 
   \midrule $-$2'241.99 &  &  &  \\ 
   \\[-2ex]  \bottomrule
\end{tabular}
}

\end{center}

\begin{figure}[t!]

\begin{Schunk}

{\centering \includegraphics{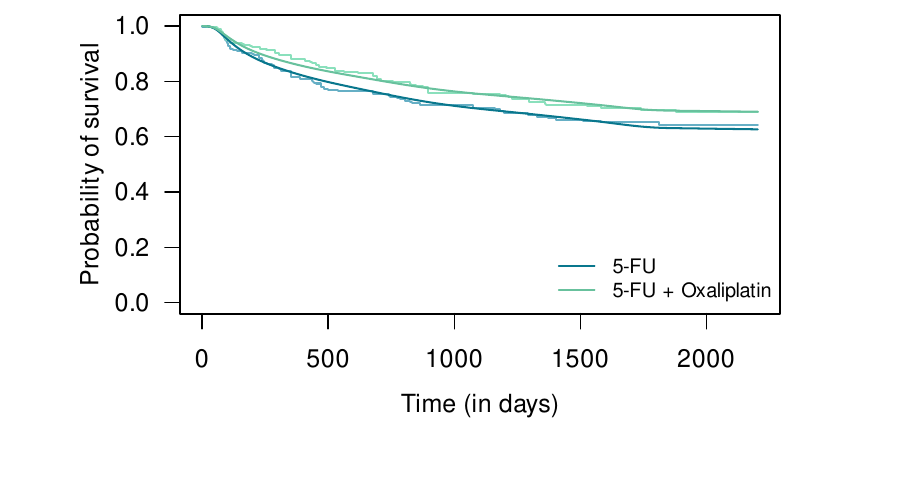} 

}

\end{Schunk}

\caption{Mixed-effects proportional hazards model.  Marginal survivor curves
obtained from integrating out the random effects from the model, along with
the non-parametric Turnbull estimates.}
\label{fig:COXME}
\end{figure}

The model provides an estimate of the log-hazard ratio $\eshiftparm$, which
is conditional on the unobserved random effects.  The estimated hazard ratio
of $\exp(\hat\eshiftparm) = 0.790$ and corresponding 95\%-confidence intervals
indicate prolonged
disease-free survival time in the experimental arm. The estimated variance of the random effect 
$R$ is relatively small, with
$\hat{\tau}^2 = 0.071$.
We can further examine
the corresponding marginal estimates of the survivor curves or related
measures by integrating out the random effects (for more details,
see \citet{Tamasi_Hothorn_2021}).  The corresponding marginal survivor curves for
patients from all blocks are depicted in Figure~\ref{fig:COXME}.

The estimated mixed-effects proportional hazards model using \cmd{CoxphME}
translates the analysis conducted in \citet{Roedel_Graeven_Fietkau_2015} into
the smooth transformation model framework.  The aim of the primary analysis of
\citet{Roedel_Graeven_Fietkau_2015} was to fit a Cox model for clustered
observations estimating the treatment effect and corresponding
standard errors.  However, at the time of the primary analysis it was not
feasible to estimate the mixed-effects Cox model while accounting for
interval-censoring.  Fortunately, here the discrepancies
between the reported results from the model ignoring interval-censoring and
the fitted one, accounting for it, are practically negligible.

To obtain a marginal hazard ratio we can estimate the model that captures the
dependence within clusters using a joint multivariate normal distribution
(Section~\ref{sec:MCOX}), which can be fitted using the \cmd{mtram}
function from the \pkg{tram} package.  Estimation is straightforward for
completely exact or interval-censored outcomes within a cluster.  Since
\var{iDFS} comprises a mix of different types of outcomes (within each
cluster, event times can be either all exact or all censored, see Formulae
2.5 and 2.6 of \citet{Barbanti_Hothorn_2023}), we handle exact
event times by treating them as interval-censored, accomplished by adding a
4-day window (stored in the object \code{iDFS2} of class `\code{Surv}', 
see Section~5 of the \code{mtram} package vignette by \citet{Barbanti_Hothorn_mtram} for details).

\begin{Schunk}
\begin{Sinput}
R> mtram(Coxph(iDFS2 ~ randarm, data = CAOsurv),
+    formula = ~ (1 | Block), data = CAOsurv)
\end{Sinput}
\end{Schunk}

\begin{center}

\scalebox{0.8}{
\begin{tabular}{rrrr}
  \toprule
Coefficient & Estimate & Std. Error & 95\%-Wald CI \\ 
  \midrule
$\eshiftparm$ & $ -0.235 $ & 0.107 & $-0.446$ to $-0.025$ \\ 
  $\mparm$ & $ 0.182 $ & 0.117 & $-0.047$ to $\phantom{-}0.411$ \\ 
   \\[-1.2ex] \toprule Log-Likelihood &   &   &   \\ 
   \midrule $-$2'047.82 &  &  &  \\ 
   \\[-2ex]  \bottomrule
\end{tabular}
}

\end{center}

The corresponding estimate of the marginal hazard ratio is
$\exp(\nicefrac{\hat\eshiftparm}{\sqrt{\hat\mparm^2 + 1}}) =
0.794$ with empirical 95\%-confidence
intervals from $0.647$ to $0.978$.  The results indicate that the
hazards for patients in the experimental arm is reduced by
$ 0.794 $ compared to the hazards in the control arm,
regardless of the block.

\section{Personalised medicine} \label{sec:PM}

In the context of personalised medicine, our focus now turns towards modeling
heterogeneous treatment effects to capture a more individualised response to treatment.
By fitting models with log-hazard ratios that vary with age, we move beyond a
global treatment effect, to assess differences in treatment response across age groups.

\subsection{Survival models}

\subsubsection{Age-varying hazards model} \label{sec:HTECOX}

To detect varying treatment effects based on age we can employ models which
estimate an age-varying hazard ratio $\exp(\eshiftparm(\ra))$ \citep{Tamasi_2025},
\begin{eqnarray*} \label{mod:HTECOX}
\sw(\ry) &=& \Prob(\rY > \ry \mid \rW = \rw,  \rA = \ra) =
\sMEV{\h(\ry) + \eshiftparm(\ra) \rw}.
\end{eqnarray*}
This formulation aligns with the model estimated in the analysis of
\citet{Hofheinz_Arnold_Fokas_2018}.

Such models could be further extended to additionally capture variations in
baseline risks across age by including an age-dependent
log-cumulative baseline hazard function: $\log(\Haznul(\ry \mid \ra)) =
\h(\ry \mid \ra) = \basisy(\ry)^\top\parm + \beta_0(\ra)$.

\subsubsection{Tree-based age-varying hazards model} \label{sec:TRT}

Furthermore, for estimating heterogeneous treatment effects, tree-based Cox
models can also be employed
\citep{Korepanova_Seibold_Steffen_2019,Seibold_Zeileis_Hothorn_2017},
\begin{eqnarray*} \label{mod:TRT}
\sw(\ry \mid \rA = \ra) &=& 
\sMEV{\h(\ry \mid \ra) + \eshiftparm(\ra) \rw},
\end{eqnarray*}
allowing to partition both the log-cumulative baseline hazard
$\log(\Haznul(\ry \mid \ra)) = \h(\ry \mid \ra) =
\basisy(\ry)^\top\parm(\ra)$ and the treatment effect $\eshiftparm(\ra)$
with respect to different age groups.  In contrast to the model in
Section~\ref{sec:HTECOX}, here both the log-cumulative baseline hazard and the
log-hazard ratio $\eshiftparm$ depend on age, in this case via an
age-structured tree.

\subsection{Application}

The hazards model featuring an age-varying treatment effect
(Section~\ref{sec:HTECOX}) can be fitted using the \pkg{tramME} package
\citep{pkg:tramME}.
\begin{Schunk}
\begin{Sinput}
R> CoxphME(iDFS ~ randarm + s(age, by = as.ordered(randarm),
+      fx = TRUE, k = 6), data = CAOsurv)
\end{Sinput}
\end{Schunk}

The model estimates a global treatment effect and an
additional smooth effect for age in the experimental arm, specified as an
unpenalized term (using \code{fx = TRUE}) to match the approach used in
\citet{Hofheinz_Arnold_Fokas_2018}. From the model terms, one can compute an
age-varying hazard ratio $\exp(\eshiftparm(\ra))$.

\begin{figure}[t!]
\centering
\begin{Schunk}

{\centering \includegraphics{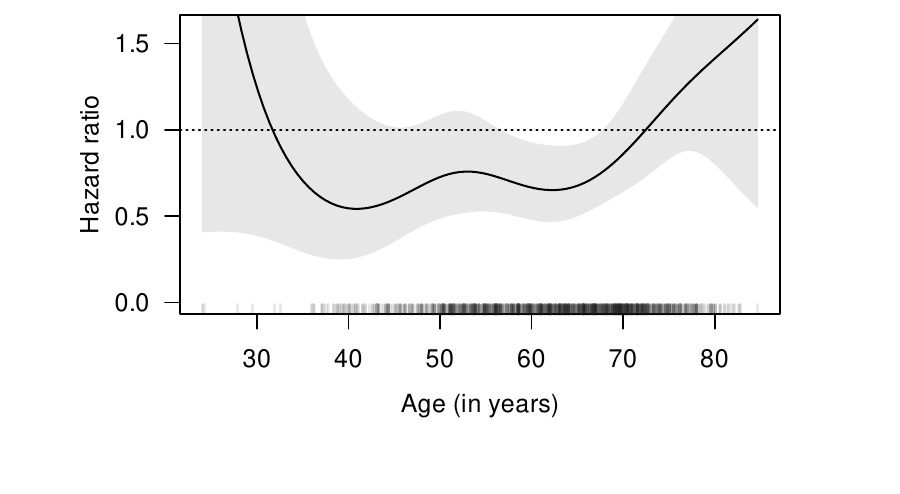} 

}

\end{Schunk}
\caption{Age-varying hazards model. Hazard ratio and corresponding 95\%-confidence interval estimated by the model shown
along age. The log-likelihood of the corresponding model is $-$2'237.63.}
\label{fig:HTECOX}
\end{figure}

The estimated age-varying hazard ratio curve, shown in Figure~\ref{fig:HTECOX},
indicates that the experimental treatment is more effective than the control
treatment for patients aged $40-70$ years, while for older patients the control
treatment reduces the hazard compared to the experimental treatment.
The corresponding 95\%-confidence interval, however, is notably wide and mostly
overlaps with a hazard ratio of 1.

Fitting a model partitioning the log-cumulative baseline hazards and
treatment effect by age, a survival tree (Section~\ref{sec:TRT}) can be estimated using the \pkg{trtf} package \citep{pkg:trtf,Hothorn_Zeileis_2017}.
\begin{Schunk}
\begin{Sinput}
R> install.packages("trtf")
R> library("trtf")
\end{Sinput}
\end{Schunk}

\begin{Schunk}
\begin{Sinput}
R> trafotree(Coxph(iDFS ~ randarm, data = CAOsurv),
+    formula = iDFS ~ randarm | age, data = CAOsurv,
+    control = ctree_control(teststat = "maximum", alpha = .1,
+      minbucket = 40))
\end{Sinput}
\end{Schunk}

\begin{figure}[t!]
\begin{Schunk}

{\centering \includegraphics{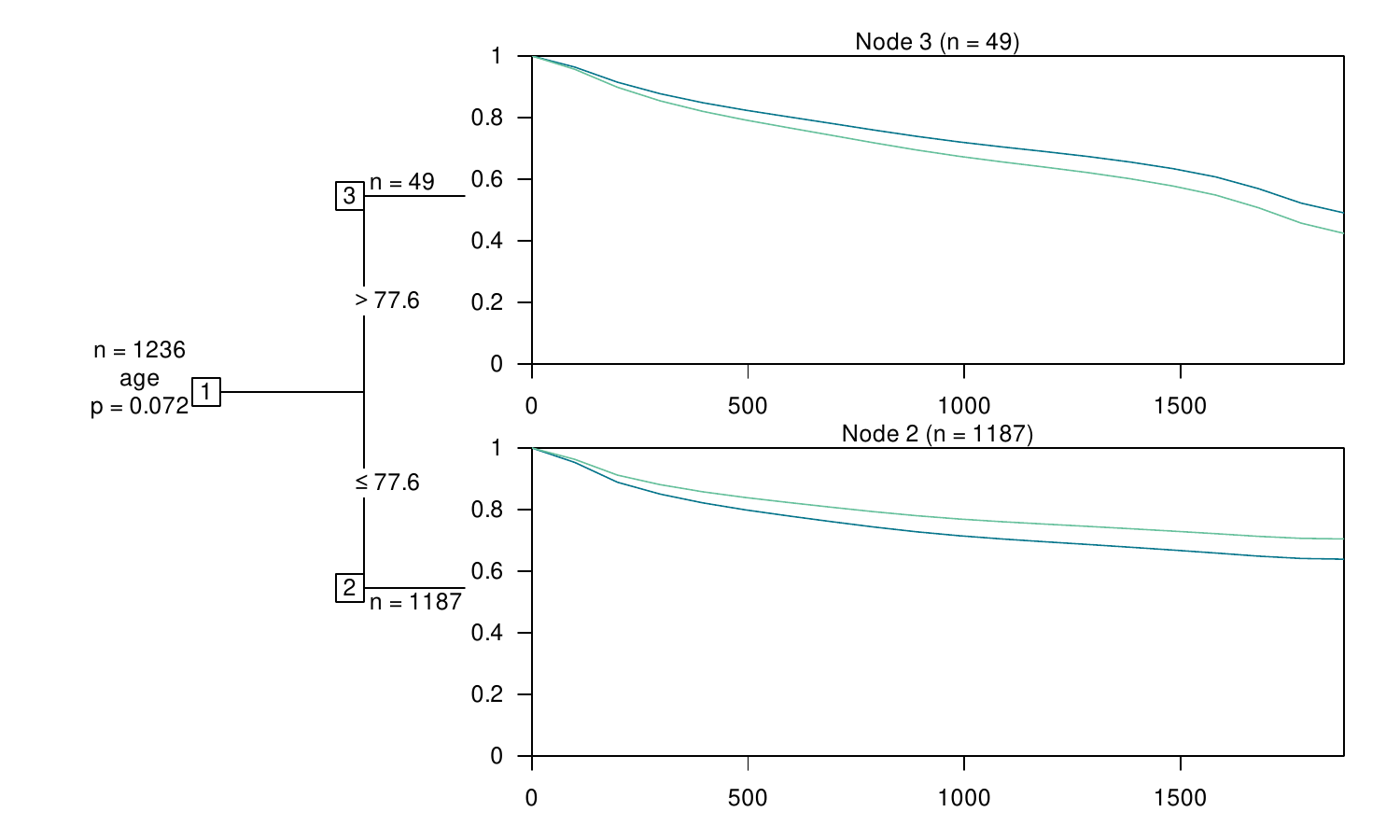} 

}

\end{Schunk}
\caption{Tree-based age-varying hazards model.
Survival tree depicting the estimated survivor curves of the age-groups
corresponding to the terminal nodes of the partitioned hazards model. The corresponding 
in-sample log-likelihood is $-$2'236.21.}
\label{fig:TRT}
\end{figure}

The survivor functions corresponding to the terminal nodes of the estimated
tree are shown in Figure~\ref{fig:TRT}.  The results again indicate that the
experimental treatment is more effective for younger patients, while the
control treatment is more effective for older patients. This result is also
in line with the one previously obtained by \citet{Hofheinz_Arnold_Fokas_2018}.

\section{Other extensions}\label{sec:ext}

\subsection{Frailty proportional hazards model}

In cases where the assumption of a homogeneous study population falls short,
frailty models offer a valuable alternative.  These models account for
unobserved heterogeneity in scenarios where the study population comprises
individuals with varying baseline risks \citep{Balan_Putter_2020}.

To handle such scenarios in the framework of smooth transformation models,
the approach discussed by \citet{McLain_Ghosh_2013} can be employed.  The
corresponding frailty proportional hazards model introduces an unobservable
multiplicative frailty effect $\rF$ \deleted{$= \rf$} on the hazard, with
corresponding conditional survivor function
\begin{eqnarray*} \label{mod:frailty}
\sw(\ry \mid \rF = \rf) &=& 
\Prob(\rY > \ry \mid \rW = \rw, \rF = \rf) = \sMEV{\h(\ry) + \log(\rf) + \eshiftparm \rw}.
\end{eqnarray*}
The model implies that the proportional hazards assumption, 
$\nicefrac{\Hazone(\ry \mid \rf)}{\Haznul(\ry \mid \rf)} =
\nicefrac{\hazone(\ry \mid  \rf)}{\haznul(\ry \mid\rf)} = \eshiftparm$,
holds conditional
on frailty $\rF = \rf$.  The frailty $\rF$ specifies a latent random
term, assumed to have a certain non-negative distribution, such as the
gamma, inverse Gaussian or positive stable distribution
\citep{Hougaard_1986}, which, for identifiability, is scaled such that
$\Ex(\rF) = 1$. 
%
%
The proportional hazards model with gamma frailty can be fitted in
\pkg{tram}, using the \cmd{Coxph} function specifying the frailty
distribution with \code{frailty = "Gamma"}.
%

%
%

\subsection{Non-parametric likelihood}

In this tutorial, we have primarily focused on the implementation of smooth
parametrisation for the log-cumulative baseline hazard function $\h$. 
Nevertheless, it is important to highlight that researchers also have the
option to utilize the discussed models that incorporate a non-parametric
version of the transformation function $\h$ in package \pkg{tram}, should they wish to do so. 
The corresponding non-parametric transformation function $\h$ is specified 
in terms of $\basisy(\ry_k)^\top \parm = \eparm_k$, where a parameter $\eparm_k$
is assigned to each distinct event time $\ry_k$ with $k = 1, \dots, K - 1$.
A head-to-head comparison of the smooth parametrisation
and the non-parametric version can be found in \citet{Yuqi_Hothorn_Li_2020}.

\subsection{Link function}

Undoubtedly, the proportional hazards model stands as a cornerstone in
survival analysis, prominently emerging from specifying the complementary
log-log link (the cumulative distribution function of the standard minimum
extreme value distribution), wherein $\h$ parametrises the log-cumulative baseline
hazard function.  Nevertheless, it is worth noting that researchers have
the option to explore other link functions for all the models shown
above, such as the logit link (as also discussed in detail in
\citet{Royston_Parmar_2002}), or the probit or log-log link.

For example, specifying a flexible proportional odds model ``only'' requires to change the link
function from complementary log-log to logit; such a model can be estimated via
\begin{Schunk}
\begin{Sinput}
R> Colr(iDFS ~ randarm, data = CAOsurv)
\end{Sinput}
\end{Schunk}

\begin{center}

\scalebox{0.8}{
\begin{tabular}{rrrr}
  \toprule
Coefficient & Estimate & Std. Error & 95\%-Score CI \\ 
  \midrule
$\eshiftparm$ & $ -0.291 $ & 0.124 & $-0.534$ to $-0.047$ \\ 
   \\[-1.2ex] \toprule Log-Likelihood & Likelihood Ratio Test & Score Test & Permutation Score Test \\ 
   \midrule $-$2'242.06 & $p$ = 0.019 & $p$ = 0.019 & $p$ = 0.019 \\ 
   \\[-2ex]  \bottomrule
\end{tabular}
}

\end{center}

This inherent versatility of link functions facilitates to construct
alternative models, including mean or odds models, by specifying a probit or 
logit link respectively.  These 
models are well known from the class of accelerated failure time models
(with log-linear $\h$), but extend seamlessly to the more flexible
framework of smooth transformation models.  Moreover, by selecting the
log-log link, it is possible to define a reverse time hazards model.  For a
comprehensive overview, see Table~1 of \citet{Hothorn_Moest_Buehlmann_2017}.

\subsection{Covariate adjustment}

While the models above focus on estimating the treatment effect,
they can naturally extend to incorporate further covariates $\rx$.
For example, in the time-varying hazards model (Section~\ref{sec:TCOX}),
age can be incorporated into the linear predictor as follows:
\begin{Schunk}
\begin{Sinput}
R> Coxph(iDFS | randarm ~ age, data = CAOsurv)
\end{Sinput}
\end{Schunk}
Additionally, penalised covariate effects can be estimated by maximizing the
$L_1$- or $L_2$-penalised log-likelihood using the \pkg{tramnet} package
\citep{pkg:tramnet,Kook_Hothorn_2021}.

Moreover, conditional transformation models \citep{Hothorn_Kneib_Buehlmann_2014},
which accommodate complex, non-linear covariate effects, can be estimated using
package \pkg{tbm} \citep{pkg:tbm, Hothorn_2020}.

\subsection{Sample size estimation and simulation}

The framework of smooth transformation models can also be valuable for
researchers involved in designing new studies.  Simulating from the
illustrated models (using \cmd{simulate}) offers a straightforward approach
for tasks such as sample size estimation. Because the transformation
function $\h$ is smooth, it is relatively simple to invert this function
numerically, such that it becomes possible to apply probability integral
transforms for generating new event times from a fitted model analogously to
\citet{Bender_Augustin_Blettner_2005}. 
As an example, we might want to generate data for a
future trial where 5-FU overall survival is improved by some innovative
therapy. We start with fitting a Weibull model to overall survival,
conditional on treatment $\rw$ and age.
\begin{Schunk}
\begin{Sinput}
R> m <- as.mlt(Survreg(OS ~ randarm + age, data = CAOsurv, 
+    dist = "weibull", support = c(.1, 80 * 365)))
\end{Sinput}
\end{Schunk}
We simulate new survival times $\rY$ from this conditional distribution
for study participants with normally distributed age in a balanced trial,
with the covariate values stored in a data frame called \code{nd}.
A useful feature in \pkg{tram} is the ability to change model coefficients
on the fly. Here, we change the log-hazard ratio to $0.25$ and simulate from
this altered Weibull model:
\begin{Schunk}
\begin{Sinput}
R> cf <- coef(m)
R> cf["randarm5-FU + Oxaliplatin"] <- .25
R> coef(m) <- cf
R> nd$T <- as.Surv(simulate(m, newdata = nd, K = 1000))
\end{Sinput}
\end{Schunk}
In addition, we simulate censoring times $\rC \indep \rY \mid \rW = \rw,  \rA = \ra$ such that $80\%$ of observations
will be right-censored (with probabilistic index
\deleted{$\Prob(\rY > \rC \mid \rW = \rw,  \rA = \ra) = \logit(1.386)$}\added{
$\Prob(\rY > \rC \mid \rW = \rw,  \rA = \ra) = 0.8 = \logit^{-1}(1.386)$}
\citep{Sewak_Hothorn_2023})
\begin{Schunk}
\begin{Sinput}
R> cf["(Intercept)"] <- cf["(Intercept)"] + qlogis(.8)
R> coef(m) <- cf
R> nd$C <- as.Surv(simulate(m, newdata = nd, K = 1000))
\end{Sinput}
\end{Schunk}
and finally compute the potentially right-censored event times for model
re-fitting:
\begin{Sinput}
R> nd$nOS <- with(nd, Surv(time = pmin(T[, "time"], C[, "time"]),
+    event = T[,"time"] < C[,"time"]))
\end{Sinput}

\section{Discussion}

Motivated by the complexities researchers face when navigating various
software implementations for survival analysis, we outline the potential of
leveraging smooth transformation models
\citep{Hothorn_Moest_Buehlmann_2017} in \proglang{R}.
Together with related add-on packages such as
\pkg{tramME} \citep{pkg:tramME} and \pkg{trtf} \citep{pkg:trtf}, the \pkg{tram} package
provides a unified maximum likelihood framework that seamlessly extends
classical survival models to accommodate more complex scenarios, offering a
versatile toolkit for survival analysis. 

Throughout this tutorial, we present practical strategies for addressing prominent
challenges in survival analysis in \proglang{R}, including interval-censored
outcomes, non-proportional or crossing hazards, dependent censoring,
clustered observations, and heterogeneous treatment effects.  The comparative
overview of implementations in Supplementary Material~\ref{sec:supp} serves
as a validation tool, allowing researchers to compare across multiple
packages.

The frameworks' modular architecture further allows
individual model components to be combined --- for example, covariate-dependent hazard
ratios can be paired with random effects using \pkg{tramME}.
The framework also extends to multiple event times per subject,
allowing for the estimation of multivariate survival models via the
\cmd{Mmlt} function \citep{Klein_Hothorn_Barbanti_2020}.
This flexibility provides users with a unified toolkit to seamlessly address a wide spectrum of
complex survival analysis tasks in \proglang{R}.

The implemented framework also provides the foundation for interesting extensions.
For example, the model in Section~\ref{sec:SCOX} could be adapted to handle cured
patients, as a fully parametric version of the semi-parametric cure mixture
model proposed by \citet{Xie_Huang_Li_2022}.
In the context of covariate adjustment, extending
the implementation to collapsible models similar to \citet{Crowther_Royston_Clements_2022} could be explored. 
Additionally, alternative strategies such as marginalising the hazard ratio, as
suggested by \citet{Rhian_2021}, could also be explored further.

\section*{Acknowledgements}
This work was supported by the Swiss National Science
Foundation [grant number 200021\_219384].

\putbib
\end{bibunit}

\newpage
\begin{appendices}
\begin{bibunit}

\section[Comparative overview of R implementations]{Comparative overview of \proglang{R} implementations}\label{sec:supp}

\definecolor{darkgray}{rgb}{0.66, 0.66, 0.66}


In the following, we compare the implementations of the models from the \pkg{tram}
\citep{pkg:tram} and \pkg{tramME} \citep{pkg:tramME} packages shown in the tutorial, with
alternative models available in various established \proglang{R}~packages
from CRAN.

We fit the  models to the primary endpoint of disease-free survival, which
comprises a mixture of exact times and right- and interval-censored event times (encoded in
\var{iDFS}, an object of class `\code{Surv}').  In order to further compare the models
with other implementations that cannot handle
interval-censored outcomes, we treat the interval-censored observations as
if they were observed exactly (encoded in \var{DFS}, an object of class `\code{Surv}').

We contrast treatment effect estimates (Estimate) and corresponding standard
errors (Std.~Error) estimated
by the fitted models.  It is important to note the difference in
interpretation of the estimates (Interpretation).  Additionally, we provide
the in-sample log-likelihood (logLik) of the fitted models, with penalised or
semi-parametric/partial likelihoods highlighted in grey. 

\subsection{Weibull models}

The \pkg{survival} package \citep{pkg:survival}, the \pkg{icenReg}
package \citep{pkg:icenReg,pkg:icenReg:JSS} and the \pkg{flexsurv} package \citep{pkg:flexsurv}
provide alternative implementations of Weibull
models.  Both the \pkg{survival} and \pkg{icenReg} package (specifying \code{model =
"aft"}) implement accelerated failure time Weibull models, where the effect
can be interpreted as log-acceleration factor (log-AF).  An alternative parametrisation of such models is in 
terms of proportional hazards Weibull models, estimating
log-hazard ratios (log-HRs), instead.  This is how the
model, fitted by \cmd{Survreg}, is implemented in the \pkg{tram} package
\citep{pkg:tram}, which can be directly compared to the analogous
parametrisation in the \pkg{icenReg} (specifying \code{model = "ph"}) and the \pkg{flexsurv} package
(with \code{dist = "weibullPH"}).  All Weibull
models can handle interval-censoring, owing to the parametric nature of the
models.

The Weibull models can be fitted to the \emph{interval-censored outcomes} as follows:
\begin{Schunk}
\begin{Sinput}
R> tram::Survreg(iDFS ~ randarm, data = CAOsurv, dist = "weibull")
R> icenReg::ic_par(iDFS ~ randarm, data = CAOsurv, dist = "weibull",
+    model = "ph")
R> flexsurv::flexsurvreg(iDFS ~ randarm, data = CAOsurv,
+    dist = "weibullPH")
R> survival::survreg(iDFS ~ randarm, data = CAOsurv, dist = "weibull")
R> icenReg::ic_par(iDFS ~ randarm, data = CAOsurv, dist = "weibull",
+    model = "aft") 
\end{Sinput}
\end{Schunk}

\begin{center}
\scalebox{0.8}{
\begin{tabular}{lllrrr}
  \toprule
Function & Package & Interpretation & Estimate & Std. Error & logLik \\ 
  \midrule
\code{Survreg} & \pkg{tram} & log-HR & $ -0.229 $ & $ 0.106 $ & $-$2'281.17 \\ 
  \code{ic\_par} & \pkg{icenReg} & log-HR & $ -0.229 $ & $ 0.106 $ & $-$2'281.17 \\ 
  \code{flexsurvreg} & \pkg{flexsurv} & log-HR & $ -0.229 $ & $ 0.106 $ & $-$2'281.17 \\ 
  \code{survreg} & \pkg{survival} & log-AF & $ 0.312 $ & $ 0.146 $ & $-$2'281.17 \\ 
  \code{ic\_par} & \pkg{icenReg} & log-AF & $ 0.312 $ & $ 0.146 $ & $-$2'281.17 \\ 
   \bottomrule
\end{tabular}
}

\end{center}

As expected, all packages provide equivalent model fits.

\subsection{Flexible proportional hazards models}

Flexible versions of the proportional hazards model are implemented in
several packages, of which the following accommodate interval-censored
outcomes.
The \pkg{rstpm2} package
\citep{pkg:rstpm2,Liu_Pawitan_Clements_2016} and the \pkg{flexsurv} package 
provide parametric versions of the model by using splines (analogously to the
approach discussed by \citet{Royston_Parmar_2002}). We set \code{k = 3} for the number 
of knots in the spline for \cmd{flexsurvspline} from the \pkg{flexsurv} package
Alternatively, the \pkg{icenReg} package \citep{pkg:icenReg} provides a
semi-parametric implementation of the model that can handle interval-censoring.

The corresponding models can be fitted to the \emph{interval-censored outcomes} as follows:

\begin{Schunk}
\begin{Sinput}
R> tram::Coxph(iDFS ~ randarm, data = CAOsurv)
R> rstpm2::stpm2(Surv(time = iDFStime, time2 = iDFStime2,
+    event = iDFSevent, type = "interval") ~ randarm, data = CAOsurv)
R> flexsurv::flexsurvspline(iDFS ~ randarm, data = CAOsurv, k = 3)
R> icenReg::ic_sp(iDFS ~ randarm, data = CAOsurv, model = "ph")
\end{Sinput}
\end{Schunk}

\begin{center}
\scalebox{0.8}{
\begin{tabular}{lllrrr}
  \toprule
Function & Package & Interpretation & Estimate & Std. Error & logLik \\ 
  \midrule
\code{Coxph} & \pkg{tram} & log-HR & $ -0.231 $ & $ 0.107 $ & $-$2'242.25 \\ 
  \code{stpm2} & \pkg{rstpm2} & log-HR & $ -0.232 $ & $ 0.107 $ & $-$2'250.48 \\ 
  \code{flexsurvspline} & \pkg{flexsurv} & log-HR & $ -0.231 $ & $ 0.106 $ & $-$2'254.34 \\ 
  \code{ic\_sp} & \pkg{icenReg} & log-HR & $ -0.230 $ & $-$ & {\color{darkgray}$-$1'977.29} \\ 
   \bottomrule
\end{tabular}
}

\end{center}

The models fit similarly across all four packages.
Due to the fact that the computations of the standard errors of \cmd{ic\_sp} from the
\pkg{icenReg} package rely on computationally expensive bootstrap sampling,
we did not report any standard errors for this approach.  Also the log-likelihood (in
grey) of the semi-parametric model from the \pkg{icenReg} is not comparable
to the otherwise fully parametric implementations.

 The \pkg{ICsurv} package \citep{pkg:ICsurv}
could also potentially handle interval-censored event times. The
\pkg{TransModel} package \citep{pkg:TransModel,pkg:TransModel:JSS},
featuring an alternative implementation of linear transformation model,
could also serve as an interesting comparator. However, we encountered
difficulties and errors when trying to fit the model using these two packages. 

In scenarios where \emph{interval-censoring} is not taken into account, there are
several other implementations available for fitting corresponding models. 
The \cmd{coxph} function from the \pkg{survival} package provides a
semi-parametric approach for exact or right-censored observations
\citep{pkg:survival}. (Note, that again the likelihood of the fitted model is not
comparable to the other fully parametric models and thus marked in grey).
The \pkg{rms} package \citep{pkg:rms} implements a semi-parametric model, in line with
the model from package \pkg{survival}.
\begin{Schunk}
\begin{Sinput}
R> tram::Coxph(DFS ~ randarm, data = CAOsurv)
R> survival::coxph(DFS ~ randarm, data = CAOsurv)
R> rms::cph(DFS ~ randarm, data = CAOsurv)
\end{Sinput}
\end{Schunk}

\begin{center}
\scalebox{0.8}{
\begin{tabular}{lllrrr}
  \toprule
Function & Package & Interpretation & Estimate & Std. Error & logLik \\ 
  \midrule
\code{Coxph} & \pkg{tram} & log-HR & $ -0.230 $ & $ 0.106 $ & $-$3'264.89 \\ 
  \code{coxph} & \pkg{survival} & log-HR & $ -0.228 $ & $ 0.106 $ & {\color{darkgray}$-$2'430.66} \\ 
  \code{cph} & \pkg{rms} & log-HR & $ -0.228 $ & $ 0.106 $ & {\color{darkgray}$-$2'430.66} \\ 
   \bottomrule
\end{tabular}
}

\end{center}

\subsection{Stratified proportional hazards models}

For comparing stratified flexible proportional hazards models we can again
utilize the model from the \pkg{rstpm2},
which employ stratified spline-basis functions. The model can be fitted to the
\emph{interval-censored event times} as follows

\begin{Schunk}
\begin{Sinput}
R> tram::Coxph(iDFS | strat ~ randarm, data = CAOsurv)
R> rstpm2::stpm2(Surv(time = iDFStime, time2 = iDFStime2,
+      event = iDFSevent, type = "interval") ~ randarm +
+      strata(strat), data = CAOsurv)
\end{Sinput}
\end{Schunk}

\begin{center}
\scalebox{0.8}{
\begin{tabular}{lllrrr}
  \toprule
Function & Package & Interpretation & Estimate & Std. Error & logLik \\ 
  \midrule
\code{Coxph} & \pkg{tram} & log-HR & $ -0.228 $ & $ 0.107 $ & $-$2'213.94 \\ 
  \code{stpm2} & \pkg{rstpm2} & log-HR & $ -0.220 $ & $ 0.107 $ & $-$2'242.88 \\ 
   \bottomrule
\end{tabular}
}

\end{center}

The results from the two models are practically similar.

Now, \emph{ignoring interval-censoring}, we can, once again, contrast the implementation of
the semi-parametric models from the
\pkg{survival} package and the \pkg{rms} package:
\begin{Schunk}
\begin{Sinput}
R> tram::Coxph(DFS | strat ~ randarm, data = CAOsurv)
R> survival::coxph(DFS ~ randarm + strata(strat), data = CAOsurv)
R> rms::cph(DFS ~ randarm + strat(strat), data = CAOsurv)
\end{Sinput}
\end{Schunk}

\begin{center}
\scalebox{0.8}{
\begin{tabular}{lllrrr}
  \toprule
Function & Package & Interpretation & Estimate & Std. Error & logLik \\ 
  \midrule
\code{Coxph} & \pkg{tram} & log-HR & $ -0.228 $ & $ 0.107 $ & $-$3'234.58 \\ 
  \code{coxph} & \pkg{survival} & log-HR & $ -0.222 $ & $ 0.107 $ & {\color{darkgray}$-$2'089.54} \\ 
  \code{cph} & \pkg{rms} & log-HR & $ -0.222 $ & $ 0.107 $ & {\color{darkgray}$-$2'089.54} \\ 
   \bottomrule
\end{tabular}
}

\end{center}

The three model fits are practically equivalent.

We can proceed to compare the stratified version of the Weibull model, for which 
we also will ignore interval-censoring due to the 
fact that the utilised \pkg{eha} package \citep{pkg:eha} lacks support for interval-censored
data.  Additionally, it is worth highlighting that there is a distinction
from the model fitted using \cmd{survreg} from the \pkg{survival} package
\citep{pkg:survival}.  This model only stratifies the scale parameter of the
Weibull distribution, whereas the models from the \pkg{eha} package and the \pkg{tram} package 
estimate both strata-dependent scale and shape terms. 
The \code{survreg} function from the \pkg{survival} package fits an accelerated failure time
Weibull model, while the \pkg{eha} package implements a proportional hazards Weibull model,
analogously to the \cmd{Survreg} implementation from the \pkg{tram} package.
The models can be fitted to the exact event times, \emph{ignoring interval-censoring},
as follows
\begin{Schunk}
\begin{Sinput}
R> tram::Survreg(DFS | strat ~ randarm, data = CAOsurv)
R> eha::phreg(DFS ~ randarm + strata(strat), data = CAOsurv)
R> survival::survreg(DFS ~ randarm + strata(strat), data = CAOsurv)
\end{Sinput}
\end{Schunk}

\begin{center}
\scalebox{0.8}{
\begin{tabular}{lllrrr}
  \toprule
Function & Package & Interpretation & Estimate & Std. Error & logLik \\ 
  \midrule
\code{Survreg} & \pkg{tram} & log-HR & $ -0.219 $ & $ 0.107 $ & $-$3'277.35 \\ 
  \code{phreg} & \pkg{eha} & log-HR & $ -0.219 $ & $ 0.107 $ & $-$3'277.35 \\ 
  \code{survreg} & \pkg{survival} & log-AF & $ 0.274 $ & $ 0.133 $ & $-$3'280.87 \\ 
   \bottomrule
\end{tabular}
}

\end{center}

The fit of the \code{survreg} model from \pkg{survival} package is slightly different.
In contrast, the parametrisation and fits of the model from the \pkg{eha} and 
the \pkg{tram} are equivalent.

\subsection{Non-proportional hazards models}

To the best of our knowledge, there is currently no implementation available that
estimates an analogous model to the flexible 
non-proportional (location-scale) hazards model implemented in \pkg{tram}.

However we can contrast the implementation of the less-flexible Weibull model
with the \pkg{gamlss} package \citep{pkg:gamlss,pkg:gamlss:JSS}
using the \code{WEI2} distribution and the \pkg{gamlss.cens} package \citep{pkg:gamlss.cens} to account for
\emph{interval-censoring}.
\begin{Schunk}
\begin{Sinput}
R> tram::Survreg(iDFS ~ randarm | randarm, data = CAOsurv,
+    remove_intercept = FALSE)
R> gamlss::gamlss(formula = iDFS ~ randarm, sigma.fo = ~ randarm,
+    family = gamlss.cens::cens(family = "WEI2", type = "interval"),
+    data = CAOsurv[, c("iDFS", "randarm")], 
+    control = gamlss.control(n.cyc = 300, trace = FALSE))
\end{Sinput}
\end{Schunk}

Since the scale term in the \pkg{tram} package and the
\pkg{gamlss} package are parameterised differently, we only show the estimates and standard errors of the
location parameter below.
\begin{center}
\scalebox{0.8}{
\begin{tabular}{llrrr}
  \toprule
Function & Package & Estimate & Std. Error & logLik \\ 
  \midrule
\code{Survreg} & \pkg{tram} & $ -0.847 $ & $ 0.536 $ & $-$2'280.47 \\ 
  \code{gamlss} & \pkg{gamlss} & $ -0.948 $ & $ 0.542 $ & $-$2'280.53 \\ 
   \bottomrule
\end{tabular}
}

\end{center}

The two implementations provide almost equivalent model fits.

The \pkg{mpr} package \citep{pkg:mpr} also offers an implementation for a
non-proportional Weibull model, it, however, does not support interval-censored data. Thus we fit the 
models \emph{ignoring interval-censoring}. 
\begin{Schunk}
\begin{Sinput}
R> tram::Survreg(DFS ~ randarm | randarm, data = CAOsurv,
+    remove_intercept = FALSE)
R> mpr::mpr(DFS ~ list(~ randarm, ~ randarm), data = CAOsurv)
\end{Sinput}
\end{Schunk}

Again, we only show the estimates and standard errors of the location parameter below.
\begin{center}
\scalebox{0.8}{
\begin{tabular}{llrrr}
  \toprule
Function & Package & Estimate & Std. Error & logLik \\ 
  \midrule
\code{Survreg} & \pkg{tram} & $ -0.976 $ & $ 0.568 $ & $-$3'290.43 \\ 
  \code{mpr} & \pkg{mpr} & $ -0.975 $ & $ 0.567 $ & $-$3'290.43 \\ 
   \bottomrule
\end{tabular}
}

\end{center}

The two implementations also provide equivalent model fits.

\subsection{Time-varying hazards model}

We can compare the time-varying hazards model from 
the \pkg{tram} and the \pkg{flexsurv} package \citep{pkg:flexsurv} which allows to 
estimate time-varying treatment effects. We start by examining the models
for the \emph{interval-censored event times}.
\begin{Schunk}
\begin{Sinput}
R> tram::Coxph(iDFS | randarm ~ 1, data = CAOsurv)
R> flexsurv::flexsurvspline(iDFS ~ randarm + gamma1(randarm) +
+      gamma2(randarm), data = CAOsurv, k = 3)
\end{Sinput}
\end{Schunk}

%
The in-sample log-likelihood is $-$2'252.95 for the \pkg{flexsurv} model 
and $-$2'240.21 for the \pkg{tram} model. The estimated
time-varying ratios of the cumulative hazards are shown in the plot below.

\begin{Schunk}

{\centering \includegraphics{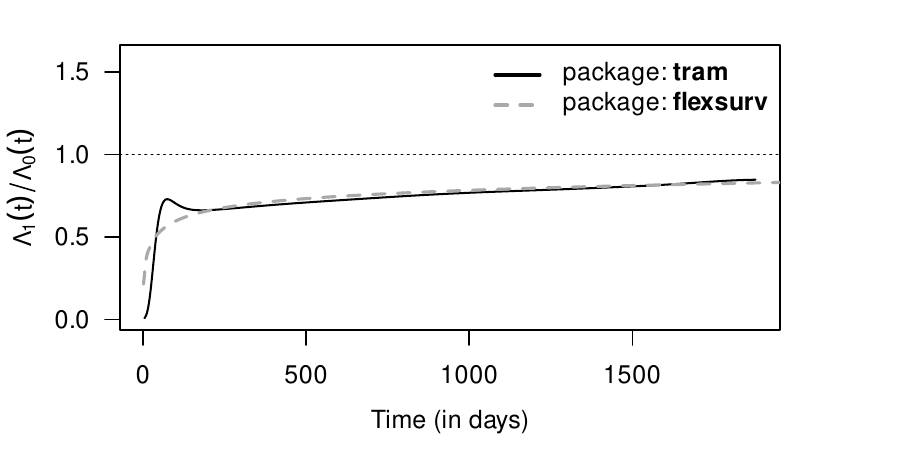} 

}

\end{Schunk}

We will now explore the same models  \emph{ignoring interval-censoring}.
\begin{Schunk}
\begin{Sinput}
R> tram::Coxph(DFS | randarm ~ 1, data = CAOsurv)
R> flexsurv::flexsurvspline(DFS ~ randarm + gamma1(randarm) +
+      gamma2(randarm), data = CAOsurv, k = 3)
\end{Sinput}
\end{Schunk}

The in-sample log-likelihood is $-$3'267.27 for the \pkg{flexsurv} model
and $-$3'262.54 for the \pkg{tram} model, with the computed time-varying ratios of the cumulative hazards shown below.
\begin{Schunk}

{\centering \includegraphics{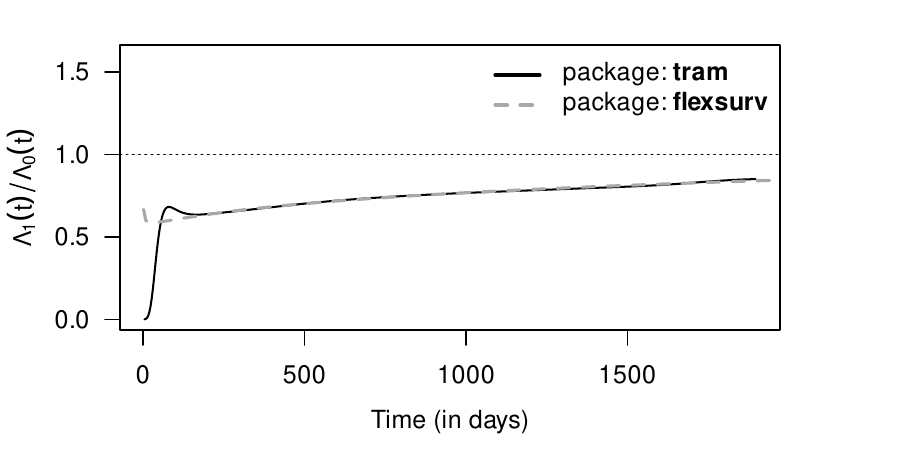} 

}

\end{Schunk}
The time-varying effects estimated from \var{DFS} show good agreement, the ratios
slightly differ when the models are estimated on the interval-censored data (\var{iDFS}).

\subsection{Mixed-effects proportional hazards models}

The implementation of a mixed-effects proportional
hazards model with flexible log-cumulative baseline hazards for
interval-censored event times is unique in the \pkg{tramME} package \citep{pkg:tramME,Tamasi_Hothorn_2021}. 
While the \pkg{rstpm2} package
also accommodates interval-censored event times,
we were not able to fit the corresponding mixed-effects model to our data.

Thus, to contrast the models with other implementations we, again,
need to \emph{ignore interval-censoring}. We can then
compare the fitted model with the
fully parametric spline-based version from 
the \pkg{rstpm2} package \citep{pkg:rstpm2,Liu_Pawitan_Clements_2017} and
the semi-parametric model estimated by the
\pkg{coxme} package \citep{pkg:coxme}, employing Gaussian random effects using a Laplace
approximation \citep{Ripatti_Palmgren_2000}.
\begin{Schunk}
\begin{Sinput}
R> tramME::CoxphME(DFS ~ randarm + (1 | Block), data = CAOsurv)
R> rstpm2::stpm2(Surv(DFStime, DFSevent) ~ randarm, data = CAOsurv,
+    cluster = "Block", RandDist = "LogN")
R> coxme::coxme(DFS ~ randarm + (1 | Block), data = CAOsurv)
\end{Sinput}
\end{Schunk}

\begin{center}
\scalebox{0.8}{
\begin{tabular}{lllrrr}
  \toprule
Function & Package & Interpretation & Estimate & Std. Error & logLik \\ 
  \midrule
\code{CoxphME} & \pkg{tramME} & log-HR & $ -0.234 $ & $ 0.107 $ & $-$3'264.66 \\ 
  \code{stpm2} & \pkg{rstpm2} & log-HR & $ -0.234 $ & $ 0.107 $ & $-$3'272.86 \\ 
  \code{coxme} & \pkg{coxme} & log-HR & $ -0.231 $ & $ 0.107 $ & {\color{darkgray}$-$2'414.48} \\ 
   \bottomrule
\end{tabular}
}

\end{center}

The fitted models from the three packages agree very well.

\subsection{Age-varying hazards model}

We can compare the age-varying hazards model from 
package \pkg{tramME} \citep{Tamasi_2025} to the implementation in the \pkg{mgcv} package
\citep{pkg:mgcv,Wood_2016} which estimates a smooth Cox model via partial likelihood optimisation.
As the model from the \pkg{mgcv} package only accommodates right-censored observations
we again fit the models \emph{ignoring interval-censoring}.

\begin{Schunk}
\begin{Sinput}
R> tramME::CoxphME(DFS ~ randarm + s(age, by = as.ordered(randarm),
+      fx = TRUE, k = 6), data = CAOsurv)
R> mgcv::gam(DFStime ~ randarm + s(age, by = as.ordered(randarm),
+      fx = TRUE, k = 6), data = CAOsurv, family = cox.ph(),
+    weights = DFSevent)
\end{Sinput}
\end{Schunk}

The in-sample log-likelihood of the model from the package \pkg{mgcv} is $-$2'426.04 (partial log-likelihood)
and $-$3'260.25 for the \pkg{tramME} model.
The estimated age-varying hazard ratios and corresponding 95\%-confidence intervals
are shown in the plot below.
\begin{Schunk}

{\centering \includegraphics{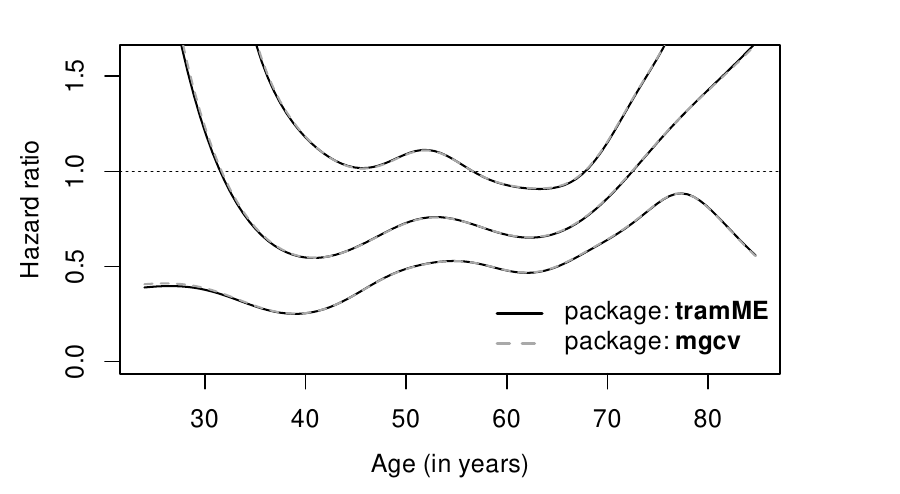} 

}

\end{Schunk}
The hazard ratio curves and confidence intervals estimated by the two packages are practically equivalent.

\subsection{Frailty proportional hazards models}

For models featuring a gamma frailty, we can contrast implementations
using a semi-parametric approach or the spline-based approach from the
\pkg{rstpm2} package \citep{Liu_Pawitan_Clements_2017}.  The \cmd{coxph} model from the \pkg{survival} package uses a semi-parametric approach and
estimates the frailty term using penalised regression \citep{Thernau_2003}.
The \pkg{frailtyEM} \citep{pkg:frailtyEM,pkg:frailtyEM:JSS} and the \pkg{frailtypack} package \citep{pkg:frailtypack,pkg:frailtypack:JSS} also feature models with
semi-parametric baseline hazards. Again we fit the models \emph{ignoring interval-censoring}.
\begin{Schunk}
\begin{Sinput}
R> tram::Coxph(DFS ~ randarm, data = CAOsurv, frailty = "Gamma")
R> rstpm2::stpm2(Surv(DFStime, DFSevent) ~ randarm, data = CAOsurv,
+    cluster = "id", RandDist = "Gamma")
R> survival::coxph(DFS ~ randarm + frailty(id, distribution = "gamma"),
+    data = CAOsurv)
R> frailtyEM::emfrail(DFS ~ randarm + cluster(id), data = CAOsurv)
R> frailtypack::frailtyPenal(DFS ~ randarm + cluster(id),
+    data = CAOsurv, RandDist = "Gamma", n.knots = 10, kappa = 1)
\end{Sinput}
\end{Schunk}

\begin{center}
\scalebox{0.8}{
\begin{tabular}{lllrrr}
  \toprule
Function & Package & Interpretation & Estimate & Std. Error & logLik \\ 
  \midrule
\code{Coxph} & \pkg{tram} & log-HR & $ -0.230 $ & $ 0.107 $ & $-$3'264.89 \\ 
  \code{stpm2} & \pkg{rstpm2} & log-HR & $ -0.685 $ & $ 0.670 $ & $-$3'264.88 \\ 
  \code{coxph} & \pkg{survival} & log-HR & $ -0.406 $ & $ 0.159 $ & {\color{darkgray}$-$1'944.22} \\ 
  \code{emfrail} & \pkg{frailtyEM} & log-HR & $ -0.384 $ & $ 0.153 $ & {\color{darkgray}$-$2'430.45} \\ 
  \code{frailtyPenal} & \pkg{frailtypack} & log-HR & $ -0.660 $ & $ 0.248 $ & {\color{darkgray}$-$3'259.82} \\ 
   \bottomrule
\end{tabular}
}

\end{center}

The fitted models vary considerably across packages.

\subsection{Flexible proportional odds models}

We can compare the fit of the flexible proportional odds model with the implementation in 
the \pkg{rstpm2} package \citep{pkg:rstpm2} and package \pkg{flexsurv} \citep{pkg:flexsurv}.
The \code{Gprop.odds} function from
package \pkg{timereg} \citep{pkg:timereg,pkg:timereg:JSS} can 
also estimate a flexible proportional odds model using the partial likelihood, thus
we again compare the models \emph{ignoring interval-censoring}.

\begin{Schunk}
\begin{Sinput}
R> tram::Colr(DFS ~ randarm, data = CAOsurv)
R> rstpm2::stpm2(Surv(DFStime, DFSevent) ~ randarm, data = CAOsurv,
+    link.type = "PO")
R> flexsurv::flexsurvspline(iDFS ~ randarm, data = CAOsurv, k = 3,
+    scale = "odds")
R> timereg::Gprop.odds(DFS ~ prop(randarm), data = CAOsurv)
\end{Sinput}
\end{Schunk}

\begin{center}
\scalebox{0.8}{
\begin{tabular}{lllrrr}
  \toprule
Function & Package & Interpretation & Estimate & Std. Error & logLik \\ 
  \midrule
\code{Colr} & \pkg{tram} & log-OR & $ -0.292 $ & $ 0.125 $ & $-$3'265.48 \\ 
  \code{stpm2} & \pkg{rstpm2} & log-OR & $ -0.294 $ & $ 0.125 $ & $-$3'272.44 \\ 
  \code{flexsurvspline} & \pkg{flexsurv} & log-OR & $ -0.294 $ & $ 0.124 $ & $-$2'247.78 \\ 
  \code{Gprop.odds} & \pkg{timereg} & log-OR & $ -0.268 $ & $ 0.125 $ &  \\ 
   \bottomrule
\end{tabular}
}

\end{center}

The fitted models are practically equivalent among the four packages.
Note, that we were not able to retrieve the in-sample log-likelihood from the model object of
the \pkg{timereg} package and thus do not report it here.

\subsection*{Computational details}
\begin{itemize}\raggedright
  \item \proglang{R} version 4.5.2 (2025-10-31), \verb|x86_64-pc-linux-gnu|
  \item Running under: \verb|Ubuntu 24.04.4 LTS|
  \item Matrix products: default
  \item BLAS:   \verb|/usr/lib/x86_64-linux-gnu/openblas-pthread/libblas.so.3|
  \item LAPACK: \verb|/usr/lib/x86_64-linux-gnu/openblas-pthread/libopenblasp-r0.3.26.so|
; \quad\ LAPACK version3.12.0
  \item Base packages: \code{base, datasets, graphics, grDevices, grid,
    methods, parallel, splines, stats, utils}
  \item Other packages: \code{ATR~0.1-1, basefun~1.2-5,
    bdsmatrix~1.3-7, boot~1.3-32, coda~0.19-4.1, coin~1.4-3,
    colorspace~2.1-2, coxme~2.2-22, doBy~4.7.1, eha~2.11.5,
    fastGHQuad~1.0.1, flexsurv~2.3.2, frailtyEM~1.0.1,
    frailtypack~3.8.0, gamlss~5.5-0, gamlss.cens~5.0-7,
    gamlss.data~6.0-7, gamlss.dist~6.1-1, Hmisc~5.2-4,
    icenReg~2.0.16, ICsurv~1.0.1, knitr~1.51, libcoin~1.0-10,
    MASS~7.3-65, mgcv~1.9-4, mlt~1.7-3, mpr~1.0.6,
    multcomp~1.4-29, mvtnorm~1.3-3, nlme~3.1-168,
    optimx~2025-4.9, parfm~2.7.8, partykit~1.2-25, Rcpp~1.1.1,
    rms~8.1-0, rstpm2~1.7.1, SparseGrid~0.8.2, survC1~1.0-3,
    survival~3.8-6, TH.data~1.1-5, timereg~2.0.7, tram~1.3-2,
    tramME~1.0.8, TransModel~2.3, trtf~0.4-3, variables~1.1-2,
    xtable~1.8-4}
  \item Loaded via a namespace (and not attached):
    \code{alabama~2025.1.0, assertthat~0.2.1, backports~1.5.0,
    base64enc~0.1-6, BB~2019.10-1, bbmle~1.0.25.1, broom~1.0.12,
    checkmate~2.3.4, cli~3.6.5, cluster~2.1.8.2,
    codetools~0.2-20, compiler~4.5.2, coneproj~1.23,
    cowplot~1.2.0, data.table~1.17.8, Deriv~4.2.0, deSolve~1.41,
    digest~0.6.39, doParallel~1.0.17, dplyr~1.1.4,
    evaluate~1.0.5, expint~0.2-1, expm~1.0-0, farver~2.1.2,
    fastmap~1.2.0, foreach~1.5.2, forecast~9.0.1, foreign~0.8-90,
    Formula~1.2-5, fracdiff~1.5-3, future~1.67.0,
    future.apply~1.20.0, generics~0.1.4, ggplot2~4.0.2,
    globals~0.19.0, glue~1.8.0, gridExtra~2.3, gtable~0.3.6,
    htmlTable~2.4.3, htmltools~0.5.8.1, htmlwidgets~1.6.4,
    inum~1.0-5, iterators~1.0.14, lattice~0.22-9, lava~1.8.2,
    lifecycle~1.0.5, listenv~0.10.0, magrittr~2.0.4,
    marqLevAlg~2.0.8, Matrix~1.7-4, matrixcalc~1.0-6,
    MatrixModels~0.5-4, matrixStats~1.5.0, microbenchmark~1.5.0,
    mnormt~2.1.2, modelr~0.1.11, modeltools~0.2-24, msm~1.8.2,
    mstate~0.3.3, muhaz~1.2.6.4, nloptr~2.2.1, nnet~7.3-20,
    numDeriv~2016.8-1.1, orthopolynom~1.0-6.1, otel~0.2.0,
    parallelly~1.46.1, pillar~1.11.1, pkgconfig~2.0.3,
    polspline~1.1.25, polynom~1.4-1, pracma~2.4.6, purrr~1.2.1,
    quadprog~1.5-8, quantreg~6.1, R6~2.6.1, rbibutils~2.4.1,
    RColorBrewer~1.1-3, Rdpack~2.6.6, reformulas~0.4.4,
    rlang~1.1.7, rmarkdown~2.30, rootSolve~1.8.2.4, rpart~4.1.24,
    rstudioapi~0.17.1, S7~0.2.1, sandwich~3.1-1, scales~1.4.0,
    sn~2.1.2, SparseM~1.84-2, statmod~1.5.1, stats4~4.5.2,
    stringi~1.8.7, stringr~1.6.0, tibble~3.3.1, tidyr~1.3.2,
    tidyselect~1.2.1, timeDate~4052.112, TMB~1.9.19, tools~4.5.2,
    urca~1.3-4, vctrs~0.7.1, withr~3.0.2, xfun~0.56, zoo~1.8-15}
\end{itemize}

\putbib
\end{bibunit}
\end{appendices}

\end{document}